\documentclass[12pt]{iopart}
\usepackage{cite}

\usepackage{graphicx}
\begin{document}
\title[]{Quasielastic hyperon production in $\bar\nu_\mu-$Nucleus interactions}
\author{M. Rafi Alam, M. Sajjad Athar, S. Chauhan and S. K. Singh}
\address{Department of Physics, Aligarh Muslim University, Aligarh - 202 002, India}
\ead{sajathar@gmail.com}
\begin{abstract}
We have studied quasielastic charged current hyperon production induced by $\bar\nu_\mu$ on free nucleon
and the nucleons bound inside the nucleus. 
 The calculations are performed for several nuclear targets like $^{12}C$,
$^{40}Ar$, $^{56}Fe$ and $^{208}Pb$ which 
are presently being used in various oscillation experiments using
accelerator neutrinos. The inputs are the hyperon-nucleon transition form
factors determined from neutrino-nucleon scattering as well as from semileptonic decays
of neutron and hyperons using SU(3) symmetry. The
calculations for the nuclear targets are done in local density approximation. The nuclear medium
effects(NME) due to Fermi motion, Pauli blocking and final state interaction(FSI) effects due to 
hyperon-nucleon scattering have been taken into account. 
\end{abstract}

\vspace{2pc}

\section{INTRODUCTION}
The present day neutrino oscillation experiments are also providing 
cross section measurements of many quasielastic
and inelastic processes induced by neutrinos and antineutrinos on nuclear targets which are being used 
in various Monte Carlo neutrino generators. 
These recent cross section measurements are available mostly for $|\Delta S| =0 $ processes in nonstrange sector. 
In strange sector the results of older work available in literature on the cross section measurements
for $|\Delta S| =0 $ and $|\Delta S|= 1$ processes are used in these neutrino generators. 
The availability of high intensity neutrino and antineutrino beams in the present generation neutrino oscillation 
experiments has opened up the possibility of experimentally studying, with better statistics, the weak production
of strange particles through $|\Delta S| =0 $ and $|\Delta S| = 1 $ processes induced by neutrinos and 
antineutrinos from nuclear targets. 
This has motivated many authors to perform new theoretical calculations of these 
processes
\cite{Singh:2006xp,Kuzmin:2008zz,Adera:2010zz,mintz,Wu:2013kla,RafiAlam:2010kf,Alam:2012zz,Alam:2013,Lalakulich:2012gm,AlvarezRuso:2012fc,Mosel:2014lja}.
In general the antineutrino - nucleus cross sections are not as well studied as the neutrino - nucleus cross sections
specifically in the region of intermediate energies  of $E_{\bar \nu} \le 2$GeV. 
In this energy region the dominant $|\Delta S| = 1 $ process is the quasielastic production of hyperons induced by 
antineutrinos, which are: 
\begin{eqnarray}\label{reaction}
\bar \nu_l + p &\longrightarrow& l^+ + \Lambda \nonumber \\
\bar \nu_l + p &\longrightarrow& l^+ + \Sigma^0 \nonumber \\
\bar \nu_l + n &\longrightarrow& l^+ + \Sigma^-;~l=e,\mu \,.
\end{eqnarray}
These processes are induced by $|\Delta S|= 1$ weak currents and are kinematically favored 
over $|\Delta S| =1 $ meson production processes. Generally, the processes like 
Eq.~\ref{reaction} are suppressed by a factor $\tan^2 \theta_C$, 
$\theta_C$ being the Cabibbo angle, as compared to the $|\Delta S| = 0$ associated production of hyperons.
 However, in the intermediate energy region of $E_{\bar \nu} < 2$ GeV, the associated production of hyperons is 
 kinematically suppressed by the phase space and the quasielastic production of hyperons shown 
 in Eq.~\ref{reaction} may be important. 
 Moreover, the experimental observation of quasielastic production of hyperons in antineutrino 
 experiments where no charged leptons are seen in the final state will give evidence of 
 Flavor Changing Neutral Current (FCNC) leading to study of physics beyond the Standard Model.  
 These reactions have been earlier studied experimentally at CERN and Serpukhov
 using Gargamelle and SKAT Bubble Chambers
 filled with heavy liquid like Freon and/or Propane and at Brookhaven National Lab(BNL)
 and Fermi National Accelerator Lab(FNAL) using Hydrogen and Hydrogen-Neon 
 targets~\cite{Eichten:1972bb,Erriquez:1978pg,Erriquez:1977tr,Ammosov:1988xb,Brunner:1989kw,Fanourakis:1980si}.
These experiments have reported 
results for the cross sections($\sigma(E)$) and $Q^2( = -q^2)$ distribution(i.e.$\frac{d\sigma}{dQ^2}$) which have large 
 uncertainties due to poor statistics. 
 It is proposed to study these reactions at MINER$\nu$A~\cite{Fields:2013zhk}, 
 MicroBooNE~\cite{Ignarra:2011yq,Karagiorgi:2012jz} and ArgoNeuT~\cite{Anderson:2011ce, Acciarri:2014isz} experiments.
 They can also be studied at other neutrino oscillation experiments at SuperK~\cite{Ashie:2005ik}, 
  MiniBooNE~\cite{AguilarArevalo:2013hm}, T2K~\cite{Abe:2011ks} and 
 NO$\nu$A~\cite{Habig:2012uva}. 
 Theoretically, these processes have been studied in past for nucleon targets either in quark model~\cite{Finjord:1975zy}
 or in  Cabibbo theory~\cite{Cabibbo:1965} using SU(3)-symmetry  of the weak hadronic 
 currents~\cite{Oset:1989ey,Ll:1972,Adler:1963nc,Egart:1963nc,Ketley:1965nc,Pais:1971ap,Marshak:1969,Bell:1996mw,Block:1964}. 
 No calculation has been made in past to study nuclear medium and final state interactions in these processes
 and it is only recently that some attempts have been made to study these effects~\cite{Singh:2006xp,Kuzmin:2008zz,Alam:2013}
 in view of the ongoing experiments~\cite{Fields:2013zhk,Ignarra:2011yq,Karagiorgi:2012jz,Anderson:2011ce,Acciarri:2014isz,Ashie:2005ik,AguilarArevalo:2013hm,Abe:2011ks,Habig:2012uva}. 

 In this paper, we present the results of a study made for quasielastic production 
 of $\Lambda$ and $\Sigma$ hyperons induced by antineutrinos on nuclear 
 targets like  $^{12}C, \; ^{40}Ar$, $^{56}Fe$ and $^{208}Pb$ 
relevant for present generation of neutrino oscillation experiments 
 being done at MINER$\nu$A\cite{Fields:2013zhk}, MicroBooNE\cite{Ignarra:2011yq, Karagiorgi:2012jz},
 ArgoNeuT\cite{Anderson:2011ce, Acciarri:2014isz}, MiniBooNE\cite{AguilarArevalo:2013hm} and T2K~\cite{Abe:2011ks}. 
We also present an estimate of nuclear medium effects(NME) and final state interaction(FSI) effect when
the reactions take place on bound nucleons in nuclei using the methods described 
in Ref.~\cite{Singh:2006xp}.
In Section-\ref{sec:qe_hyp}, we briefly review the formalism and various assumptions used to 
calculate $|\Delta S| =1$ quasielastic reactions 
and define various quantities used in this work. 
We describe the nuclear medium and final state interaction effects in Section-\ref{sec:results} and 
present the numerical results for the total scattering cross sections and the differential scattering
cross sections relevant for various experiments where measurements of quasielastic hyperon 
production may be made in near future. In Section-\ref{sec:summary}, we present a summary and future outlook of these processes 
in view of present antineutrino experiments. 

\section{QUASIELASTIC PRODUCTION OF HYPERONS}\label{sec:qe_hyp}
$|\Delta S|=1$ hyperon $(Y)$ production processes induced by muon type antineutrinos which
are presently being used in accelerator experiments are written as
\begin{eqnarray}\label{hyp-rec}
{\bar\nu_\mu}(k) + p(p)&\rightarrow \mu^+(k^\prime) + \Lambda(p^\prime) \nonumber \\ 
{\bar\nu_\mu}(k) + p(p)&\rightarrow \mu^+(k^\prime) + \Sigma^0(p^\prime)\nonumber \\ 
{\bar\nu_\mu}(k) + n(p)&\rightarrow \mu^+(k^\prime) + \Sigma^-(p^\prime) 
\end{eqnarray} 
for which the differential scattering cross section in the laboratory frame is given by
\begin{equation}
\label{crosv.eq}
d\sigma=\frac{1}{(2\pi)^2}\frac{1}{4E_{\bar \nu} M}\delta^4(k+p-k^\prime-p^\prime)
\frac{d^3k^\prime}{2E_{k^\prime}}\frac{d^3p^\prime}{2E_{p^\prime}}\sum \bar{\sum} |{\cal{M}}|^2 ,
\end{equation}
where $M$ is the nucleon mass and ${\cal{M}}$ is the transition matrix 
element given by
\begin{equation}\label{matrix1}
{\cal{M}}=\frac{G_F}{\sqrt{2}}\sin\theta_c l^{\mu}~J_\mu.
\end{equation}
In the above expression $l^{\mu}$ is the leptonic current($\bar v(k^\prime) \gamma^\mu (1+\gamma_5) v(k)$) and $J_\mu(|\Delta S|=1)$
is the matrix element of strangeness changing hadronic current defined as
\begin{equation}\label{had}
 J_{\mu} = \langle Y(p^{\prime})|V_{\mu} - A_{\mu}|N(p) \rangle ,
\end{equation}
where coupling to the leptonic current $l^\mu$ is determined as $G_{F} \sin\theta_C/\sqrt{2}$ in terms Cabibbo angle $\theta_C$ using
universality of weak interactions. In Eq.~\ref{had}, $Y(p^{\prime})$ and $N(p)$ denote the final hyperon $Y$ and initial 
nucleon N with momenta $p^{\prime}$ and $p$, respectively and
the matrix element of vector and axial vector currents are defined as
\begin{eqnarray}\label{vec}
 \langle Y(p^{\prime})|V_{\mu}|N(p) \rangle &=& {\bar{u}_Y}(p^\prime)\left[\gamma_\mu f_1(q^2)+i\sigma_{\mu\nu} 
\frac{q^\nu}{M+M_Y} f_2(q^2) \right. \nonumber\\ 
&+& \left. \frac{f_3(q^2)}{M + M_Y} q_\mu \right]u_N(p)
\end{eqnarray}
and
\begin{eqnarray}\label{axi}
 \langle Y(p^{\prime})|A_{\mu}|N(p) \rangle &=& {\bar{u}_Y}(p^\prime)\left[\gamma_\mu \gamma_5 g_1(q^2) + 
 i \sigma_{\mu\nu}\gamma_5 \frac{q^\nu}{M+M_Y} g_2(q^2) \right. \nonumber\\  
 &+& \left. \frac{g_3(q^2)} {M + M_Y} q_\mu \gamma_5 \right]u_N(p),
\end{eqnarray}
where $q^2$ is the four momentum transfer square($q^2$=$-Q^2$,$Q^2 \ge 0$) and $M_Y$ is the mass of hyperon. In defining the matrix 
elements of vector and axial vector currents in Eqs.~\ref{vec} and \ref{axi} 
we have followed the conventions used by Llewellyn Smith\cite{Ll:1972},
while other conventions also exist in literature specially for the momentum 
dependent terms involving $f_2(q^2)$, $f_3(q^2)$, $g_2(q^2)$ and
$g_3(q^2)$. This is discussed further in Section-\ref{sec:form}, 
when we determine them using symmetry properties of weak hadronic currents including SU(3)
symmetry.
\subsection{Form Factors}\label{sec:form}
The six form factors $f_i(q^2)$ and $g_i(q^2)$ ($i=1,2,3$) are determined using following assumptions about the weak vector and
axial vector currents in weak interactions.
\begin{enumerate}
\item[(a)] The assumption of T invariance implies that all the form factors $f_i(q^2)$ and $g_i(q^2)$ are real.
\item[(b)] The assumption of SU(3) symmetry of weak hadronic currents implies that the vector and axial vector 
currents have definite transformation
properties under SU(3) group of transformations. Assuming that $|\Delta S|= 0$ and $|\Delta S|= 1$ weak currents 
along with the electromagnetic
currents transform as octet representation under SU(3), determine various couplings of these currents to initial and final baryons 
using SU(3) Clebsch-Gordan coefficients corresponding to the decomposition 
\begin{equation}
 8  \otimes 8 = 1  \oplus 8^{S}  \oplus 8^{A}  \oplus 10 \oplus 10 \oplus 27 .
\end{equation}
Since initial and final baryons also belong to octet representation, each form factor  $f_i(q^2)$ ($g_i(q^2)$) occurring in the matrix element
of vector(axial vector) current is written in terms of two functions $D(q^2)$ and $F(q^2)$ corresponding to symmetric octet($8^{S}$) 
and antisymmetric octet($8^{A}$) couplings of octets of vector(axial vector) currents. Specifically we write
\begin{eqnarray}\label{coefficients}
 f_i(q^2) = a F_{i}^{V}(q^2) + b D_{i}^{V}(q^2)\nonumber\\
 g_i(q^2) = a F_{i}^{A}(q^2) + b D_{i}^{A}(q^2),
\end{eqnarray}
where a and b are SU(3) Clebsch-Gordan coefficients given in Table-\ref{tab1}. 
We find from the values of a and b given in Table-\ref{tab1} that 
\begin{equation}
\frac{d\sigma}{dq^2}(\bar \nu_{\mu} p \rightarrow \mu^{+} \Sigma^{0})  = 
 \frac{1}{2} \frac{d\sigma}{dq^2}(\bar \nu_{\mu} n \rightarrow \mu^{+} \Sigma^{-}) \, .
\end{equation}
\item[(c)] The assumption of SU(3) symmetry and G invariance together imply the absence of 
second class currents~\cite{Weinberg:1958pr} leading to
\begin{equation}
 f_{3}(q^{2})  = 0.
\end{equation}
\item[(d)] The assumption of Conserved Vector Current and SU(3) symmetry 
implies $f_{3}(q^{2}) = 0$ and leads to the determination of 
$f_{1}(q^{2})$ and $f_{2}(q^{2})$ in terms of electromagnetic 
form factors of nucleons $f_{1}^{N}(q^{2})$ and $f_{2}^{N}(q^{2}); (N = p, n)$.

In order to do this, we write electromagnetic current in terms of its SU(3) content as 
\begin{equation}
 V_{\mu}^{em} = V_{\mu}^{3} + \frac{1}{\sqrt{3}} V_{\mu}^{8},
\end{equation}
where the superscript 3 and 8 denote SU(3) indices. We define the matrix element of the
electromagnetic current between nucleon states in terms of electromagnetic form factors of nucleons 
$f_{1}^{N}(q^{2})$ and $f_{2}^{N}(q^{2})$ as 
\begin{eqnarray}\label{EM_current}
 \langle N(p^{\prime}) | V_{\mu}^{em} | N(p) \rangle &=& \bar u_{N}(p^{\prime})\left[\gamma_\mu f_{1}^{N}(q^2)+i\sigma_{\mu\nu} 
\frac{q^\nu}{2M} f_{2}^{N}(q^2)\right]u_N(p).
\end{eqnarray}
Evaluating Eq.~\ref{EM_current} between nucleon states, we find 
\begin{eqnarray}\left.
\begin{array}{l}
f_{i}^{p}(q^2) = F_{i}^{V}(q^2) + \frac{1}{3} D_{i}^{V}(q^2) \\
 f_{i}^{n}(q^2) = - \frac{2}{3} D_{i}^{V}(q^2); 
\end{array}\right\} \quad i = 1,2   
\end{eqnarray}
which determines the functions $D_{i}^{V}(q^2)$ and $F_{i}^{V}(q^2)$ corresponding to the non-vanishing form factors $f_{i}(q^2)$ $(i = 1, 2)$ in terms
of the electromagnetic form factors of nucleons. This along with the Clebsch-Gordan coefficients 
$a$ and $b$ given in Table-\ref{tab1}, completely determine all the vector form factors 
and are given in Table-\ref{tab2}. It may be noted that we have implemented the 
SU(3) symmetry at the level of form factor $f_{2}(q^{2})$ 
and not at the level of $f_{2}(q^{2})/M_{Y}$ as done in Refs.~\cite{Cabibbo:2003cu, Gaillard:1984ny} using  Cabibbo model for the 
  analysis of semileptonic decays. 

\item[(e)]  In the axial vector sector, the form factor $g_{2}(q^{2})$ vanishes due to G invariance and SU(3) symmetry. 
The contribution of $g_{3}(q^{2})$ 
is very small as it is proportional to the lepton mass in the 
matrix element and is generally neglected in the case of $|\Delta S| = 0$ reactions. 
We neglect it here in the case of $|\Delta S|= 1$ reactions. 
Thus the only non-vanishing form factor is therefore
$g_{1}(q^{2})$ which is determined in terms of the two functions $D_{1}^{A}(q^2) = D(q^2)$ and $F_{1}^{A}(q^2) = F(q^2)$. 
With the values of 
Clebsch-Gordan coefficients $a$ and $b$ given in Table-\ref{tab1}, 
we tabulate them in Table-\ref{tab2} for the reactions studied in this paper.

It should be emphasized that the determination of various vector(axial vector) form factors $f_{i}(q^{2})$($g_{i}(q^{2})$) (i = 1, 2, 3)
depends very crucially on SU(3) symmetry which is known to work
well in the analysis of semileptonic decays of hyperons provided the physical masses for hyperons are used in the analysis.
But the question of 
SU(3) symmetry in $|\Delta S|= 1$ reactions
induced by (anti)neutrinos is yet to be studied and should be investigated specially when 
there are indications of non-zero $g_{2}(q^{2})$ in the analysis 
of semileptonic hyperon decays. With antineutrino beam of high intensity,
the study of strange particle production through $|\Delta S|= 1$ reactions shall provide an
opportunity to study SU(3) breaking effects as well 
as G invariance in the strangeness sector.
\end{enumerate}
\subsection{$q^2$ dependence of form factors}\label{sec:q2dep}
\begin{enumerate}
\item[(i)] Vector form factors\\
The vector transition form factors for $p \rightarrow \Lambda$ and $p \rightarrow \Sigma^{0}$ transitions given in terms of 
the electromagnetic form factors of neutrons and protons as shown in Table-\ref{tab2}, 
are written in terms of Sach's form factors $G_{E}^{p, n}(q^2)$ and $G_{M}^{p, n}(q^2)$ as
\begin{eqnarray}\label{Vec_FF}
 f_{1}^{p, n}(q^2) &=& \frac{1}{1 - \frac{q^2}{4M^2}} \left[ G_{E}^{p, n}(q^2) - \frac{q^2}{4M^2} G_{M}^{p, n}(q^2)\right] \nonumber\\
 f_{2}^{p, n}(q^2) &=& \frac{1}{1 - \frac{q^2}{4M^2}} \left[ G_{M}^{p, n}(q^2) - G_{E}^{p, n}(q^2)\right] .
\end{eqnarray}
The Sach's form factors $G_{E}^{p, n}(q^2)$ and $G_{M}^{p, n}(q^2)$ are parameterized as 
\begin{eqnarray}
 G_{E}^{p}(q^2) &=& \left( 1 - \frac{q^2}{M_{V}^{2}}\right)^{-2}, \nonumber\\
 G_{M}^{p}(q^2) &=& (1 + \mu_{p}) G_{E}^{p}(q^2), \nonumber\\
 G_{M}^{n}(q^2) &=& \mu_{n} G_{E}^{p}(q^2), \nonumber\\
 G_{E}^{n}(q^2) &=& \frac{q^2}{4M^2} \mu_{n} G_{E}^{p}(q^2) \xi_{n};
\end{eqnarray}
The numerical values of various parameters are taken as, 
\begin{eqnarray}
\label{eq:xin_def}
 \xi_{n} &=& \frac{1}{1 - \lambda_{n} \frac{q^2}{4M^2}}, \nonumber\\
 \mu_{p} &=& 1.792847, \nonumber\\
 \mu_{n} &=& -1.913043, \nonumber\\
 M_{V} &=& 0.84 GeV  \;\; \rm{and} \;\;  \lambda_{n} = 5.6
\end{eqnarray}
\item[(ii)] Axial vector form factors\\
With $g_{2}(q^{2}) = 0$ and the contribution of $g_{3}(q^{2})$ being negligible, only $g_{1}(q^{2})$ contributes to 
the cross section for reactions considered in this paper. 
$g_{1}(q^{2})$  is described in terms of two functions $F(q^2)$ and $D(q^2)$ for all reactions in $|\Delta S|= 1$ sector. 
A priori there is  no reason to assume same $q^2$ dependence
for $F(q^2)$ and $D(q^2)$; but if we do that then all the transitions are determined in terms of one function for the $g_{1}(q^{2})$
form factor which is chosen to be $F(q^2) + D(q^2)$(=$g_{A}^{n \rightarrow p}(q^2)$)
and is determined from $|\Delta S| = 0$, neutrino (antineutrino) reactions on nucleons
and a constant term $x$ given by:
\begin{equation}\label{Eq:xdep}
  x =  \frac{F(q^2)}{F(q^2) + D(q^2)}
= \frac{F(0)}{F(0) + D(0)} 
\end{equation}
The above relation given by Eq.~\ref{Eq:xdep} is valid, if the same $q^2$ dependence is assumed for $F(q^2)$ and $D(q^2)$. For numerical calculations we take $F(0)=0.463$ and 
$D(0)=0.804$~\cite{Cabibbo:2003cu}.

With these assumptions, $g_{1}(q^{2})$ form factor for various transitions in Table-\ref{tab2} are given as 
\begin{eqnarray}
 g_{1}^{p \rightarrow \Lambda} (q^{2}) = - \sqrt{\frac{3}{2}} \frac{1 + 2x}{3} g_{A}(q^{2}) \nonumber\\
 g_{1}^{p \rightarrow \Sigma^0} (q^{2}) = \sqrt{\frac12} (1 - 2x) g_{A}(q^{2})
\end{eqnarray}
where 
\begin{eqnarray}\label{Axi_FF}
 g_{A}(q^{2}) &=& g_{A}(0) \left(1 - \frac{q^2}{M_{A}^{2}}\right)^{-2}.
\end{eqnarray}

We use  $g_{A}(0) = 1.267$~\cite{Cabibbo:2003cu} and $M_{A} = 1.03GeV$~\cite{Bernard:2001rs} for the numerical calculations.  
\begin{table}\centering
\renewcommand{\arraystretch}{2}
 \begin{tabular}{c|cc}\hline \hline
Transitions & a & b \\ \hline \hline 
p $\rightarrow \Lambda$ & $-\sqrt{\frac{3}{2}}$ & $ -\sqrt{\frac{1}{6}}$  \\ \hline 
n $\rightarrow \Sigma^{-}$ & $-1$ & 1  \\ \hline 
p $\rightarrow \Sigma^{0}$ & $-\frac{1}{\sqrt{2}}$ & $\frac{1}{\sqrt{2}}$ \\ \hline \hline 
 \end{tabular}
\caption{Values of the coefficients $a$ and $b$ of the form factors given in Eq.~\ref{coefficients}.}
\label{tab1}
\end{table}
\end{enumerate}
\subsection{Nuclear Effects}\label{sec:nuclear_effects}
When the reactions shown in Eq.~\ref{hyp-rec} take place on nucleons which are bound in the
nucleus, Fermi motion and Pauli blocking
effects of the nucleon is to be considered. In the final state the produced hyperon
is affected by the final state interactions with
the nucleons inside the nucleus through the hyperon-nucleon quasielastic and
charge exchange scattering processes. The  Fermi 
motion effect is calculated in a local Fermi Gas model and the cross section
is evaluated as a function of local Fermi momentum, $p_F(r)$ and 
integrated over the whole nucleus. In a nucleus, the neutrino scatters from a
neutron or a proton whose local density in the medium is $\rho_n(r)$ or
$\rho_p(r)$, respectively. 
The corresponding local Fermi momentum for neutrons and protons are given as 
\begin{equation}
{p_F}_n={[3\pi^2\rho_n(r)]}^{1/3};
{p_F}_p={[3\pi^2\rho_p(r)]}^{1/3}.
\end{equation}

The differential scattering cross section for the scattering of antineutrinos 
from nucleons in the nucleus is then given as 
\begin{equation}\label{diffnuc}
\frac{d\sigma}{d\Omega_ldE_l}=2{\int d^3r \int \frac{d^3p}{{(2\pi)}^3}n_N(p,r)
\left[\frac{d\sigma}{d\Omega_ldE_l}\right]_{free}}
\end{equation}
where ${(\frac{d\sigma}{d\Omega_ldE_l})}_{free}$ is the differential cross
section for free antineutrino nucleon scattering  given in Eq.~\ref{crosv.eq} and $n_N(p,r)$ is local occupation number of the 
nucleon of momentum $p$ at a radius $r$ in the nucleus, which is 1 for $p < p_{F_N}$ and 0 otherwise.
 Moreover, the initial state interaction of nucleons inside the nuclear medium leading to $2p-2h$
excitations due to nucleon correlation and the meson  exchange current effects describing physics beyond 
the independent particle description of nucleus are known to be quite important in the quasielastic  
(anti)neutrino-nucleus reactions in the $\Delta S =0 $ sector~\cite{Marteau:1999kt,Meucci:2011vd,
Amaro:2011qb,Nieves:2011pp,Martini:2009uj}. 
These effects may also play an important role in  quasielastic hyperon production but a study of these effects 
is beyond the scope of this paper and should be taken up in future.

\subsection{Final State Interactions}\label{sec:final_state_int}
The relative yields of $\Lambda$ and $\Sigma$  hyperons produced in the initial weak production processes are 
modified due to hyperon-nucleon final state interactions in the nuclear medium.   
The final state interactions induce elastic $\Lambda-N$ and $\Sigma-N$ scattering, 
charge-exchange $\Sigma-N$ scattering and inelastic $\Sigma N \rightarrow \Lambda N$ and $\Lambda N \rightarrow \Sigma N$ reactions.
In general $\Lambda N \rightarrow \Sigma N$ reactions are kinematically inhibited as compared to $\Sigma N \rightarrow \Lambda N$ 
reactions leading to an enhancement in $\Lambda$ production and depletion in $\Sigma$ production  
as a result of final state interactions.
An interesting feature of final state interactions is the production of $\Sigma^+$ through 
$\Lambda p \rightarrow \Sigma^+ n $ and $\Sigma^0 p \rightarrow \Sigma^+ n $ processes, 
as the production of $\Sigma^+$ is not allowed through the basic weak processes in $\bar \nu N $ scattering.

For the final state interaction of hyperons we have followed
Ref.~\cite{Singh:2006xp}. In this prescription an initial hyperon produced at a position 
$r$ within the nucleus interacts with a nucleon to produce a new hyperon state
within a short distance $dl$ with a probability $P = P_{Y}dl$, where
$P_Y$ is probability per unit length given by 
\[P_Y=\sigma_{Y+n \rightarrow f}(E)~\rho_{n}(r)~+~\sigma_{Y+p \rightarrow
f}(E)~\rho_{p}(r),\]
 where $f$ denotes a possible final hyperon-nucleon ($Y_f(\Sigma ~{\rm or}~
\Lambda) + N(n~or~p)$) state with energy E in the hyperon-nucleon CM system, 
$\rho_{n}(r)(\rho_{p}(r))$ is the local density of neutron(proton) in the
nucleus and $\sigma$ is the total cross section for the charged current channel
 like $Y(\Sigma ~{\rm or}~ \Lambda)~+~N(n~or~p) \rightarrow
f$~\cite{Singh:2006xp}. Now a particular channel is selected giving rise to a
hyperon $Y_f$ in the final state with the probability $P$. 
For the selected channel Pauli blocking effect is taken 
into account by first randomly selecting a nucleon in the local Fermi sea. Then
a random scattering angle is generated in the hyperon-nucleon CM system assuming
the cross sections to be isotropic.
 Using this information hyperon and nucleon momenta are calculated and Lorentz
boosted to lab frame. 
 If the nucleon in the final state has momenta above the Fermi momenta we have a
new hyperon
type($Y_f$) and/or a new direction and energy of the initial hyperon($Y_i$).
This process is continued until the hyperon gets out of the nucleus.
For numerical evaluations, we have taken the hyperon-nucleon cross sections given in 
Refs.~\cite{Alexander:1969cx,SechiZorn:1969hk,Kadyk:1971tc,Hauptman:1977hr,Eisele:1971mk,Charlton:1970bv}
and parameterized in Ref.~\cite{Singh:2006xp}.
The nuclear densities for various nuclei are taken from Refs.~\cite{GarciaRecio:1991wk,DeJager:1987qc}.

\begin{table}\centering
\renewcommand{\arraystretch}{2}
 \begin{tabular}{c|cc}\hline \hline
FF & $p \rightarrow \Sigma^0 $ & $p \rightarrow \Lambda $ \\ \hline \hline 
$f_1(q^2)$ & $\frac{-1}{\sqrt2} (f_1^p(q^2) + 2 f_1^n(q^2))$ & $ -\sqrt{\frac32} f_1^p(q^2) $  \\ \hline 
$f_2(q^2)$ & $\frac{-1}{\sqrt2} (f_2^p(q^2) + 2 f_2^n(q^2))$ & $ -\sqrt{\frac32} f_2^p(q^2) $  \\ \hline 
$g_1(q^2)$ & $\frac{1}{\sqrt2}\frac{D-F}{D+F} g_A(q^2)$ & $-\frac{D+3 F}{\sqrt{6} (D+F)} g_A(q^2)$ \\ \hline \hline 
 \end{tabular}
\caption{Form factors of Eqs.~\ref{vec} and ~\ref{axi}. 
$f_i^N(q^2), i=1,2, N=n,p$ are defined in Eq.~\ref{Vec_FF} and $g_A(q^2)$ is defined in Eq.~\ref{Axi_FF}.
The parameters $F$ and $D$ are determined from the semileptonic decays which for the present work 
are taken as 0.463 and 0.804 respectively.}
\label{tab2}
\end{table}

 \begin{figure}
 \centering
\includegraphics[height=0.26\textheight,width=0.7\textwidth]{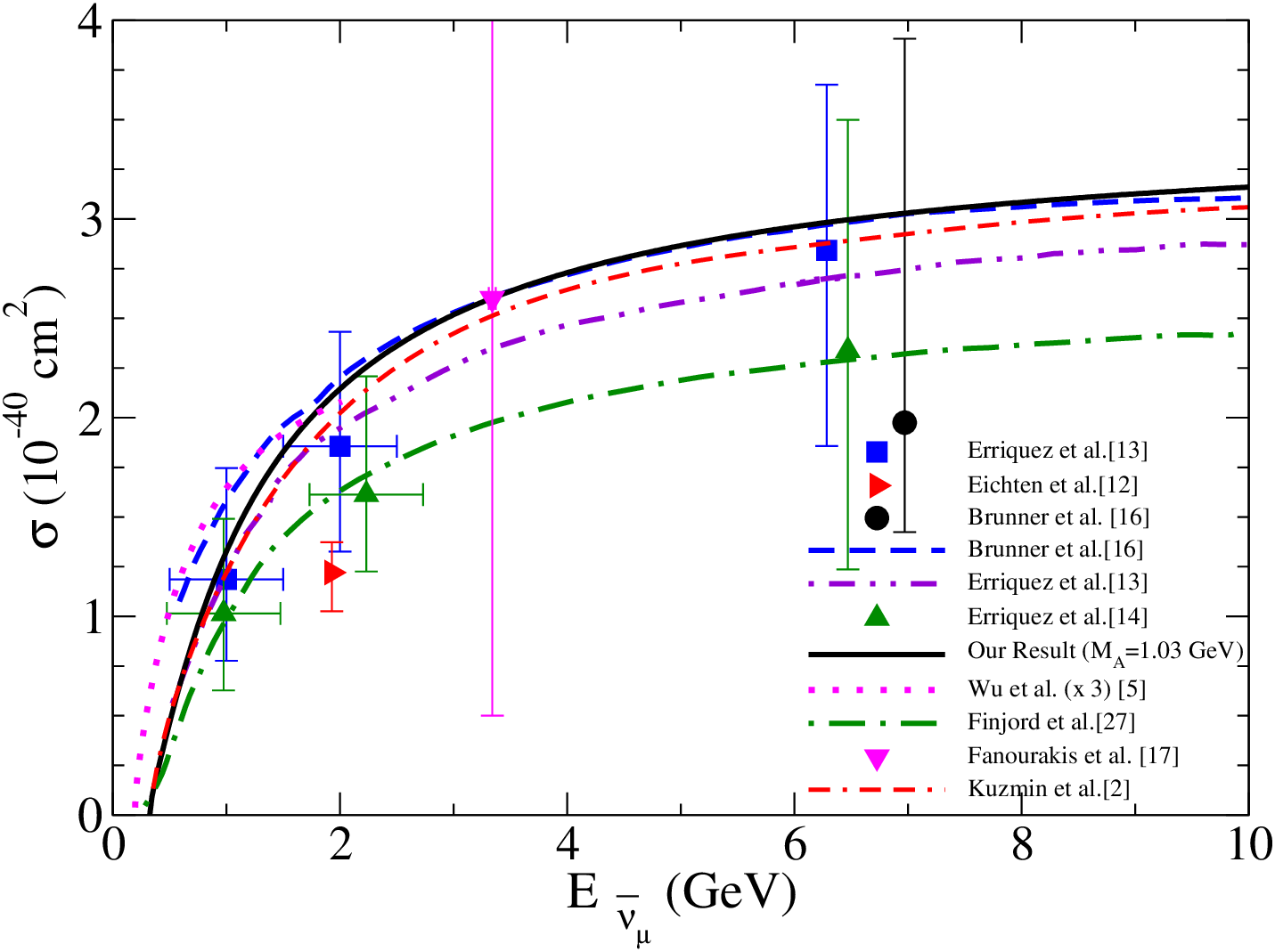}
\caption{(color online). $\sigma$ vs $E_{\bar\nu_\mu}$, for ${\bar\nu_\mu} + p \rightarrow \mu^+ 
+ \Lambda$ process. Experimental results (triangle right~\cite{Eichten:1972bb}, square~\cite{Erriquez:1978pg},
triangle up~\cite{Erriquez:1977tr}, circle~\cite{Brunner:1989kw}), 
triangle down($\sigma = 2.6^{+5.9}_{-2.1} \times 10^{-40} cm^2$)~\cite{Fanourakis:1980si}
are shown with error bars. Theoretical curves are of Kuzmin and Naumov~\cite{Kuzmin:2008zz}(double dashed-dotted line), 
Brunner et al.~\cite{Brunner:1989kw}(dashed line),  
 Erriquez et al.~\cite{Erriquez:1978pg}(dashed-double dotted line) obtained using Cabibbo theory  
 with axial vector dipole mass as 0.999GeV, 1.1 GeV and 1 GeV, respectively, while
 the results of Wu et al.~\cite{Wu:2013kla}(dotted line) and Finjord and 
Ravndal~\cite{Finjord:1975zy}(dashed dotted line) are obtained using quark model. The results of present calculation are 
shown with solid line. Notice that we have multiplied the results of Wu et al.~\cite{Wu:2013kla} by 3 to plot on the same scale.}
\label{fg:xsec_comp}
\end{figure}

\begin{figure}\centering
\includegraphics[height=0.26\textheight,width=0.7\textwidth]{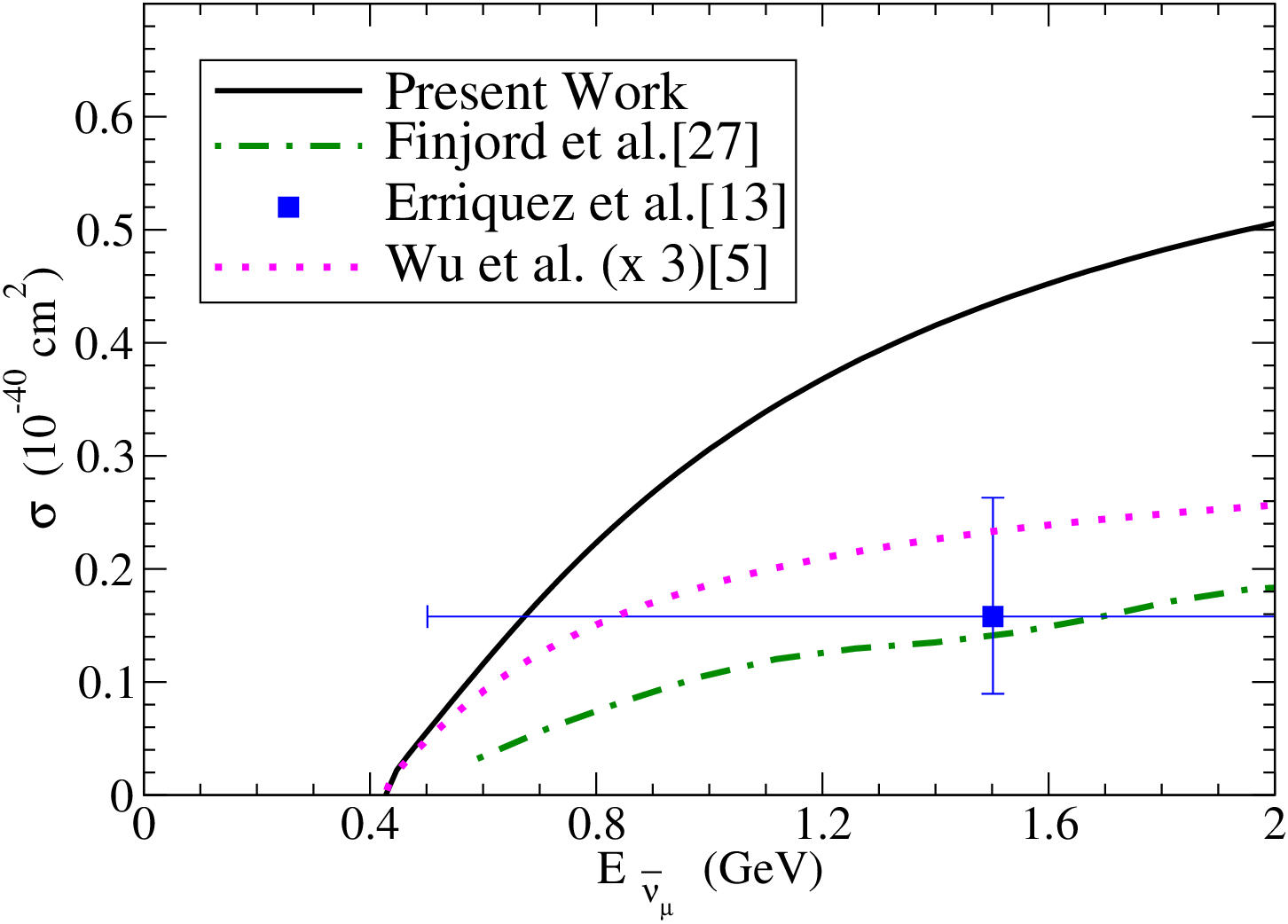}
\caption{(color online). $\sigma$ vs $E_{\bar\nu_\mu}$, for $\bar\nu_\mu + p \rightarrow \mu^+ + \Sigma^0$ process. 
Experimental points is taken from Ref.~\cite{Erriquez:1978pg}.
Present results are shown with solid line. Also the results of Wu et al.~\cite{Wu:2013kla}(dotted line) 
and  Finjord and Ravndal~\cite{Finjord:1975zy}(dashed dotted line) have been presented. 
Notice that we have multiplied the results of Wu et al.~\cite{Wu:2013kla} by 3 to plot on the same scale.}
\label{fg:xsec_p_mu_Sig}     
\end{figure}

\begin{figure}\centering
\includegraphics[height=0.26\textheight,width=0.7\textwidth]{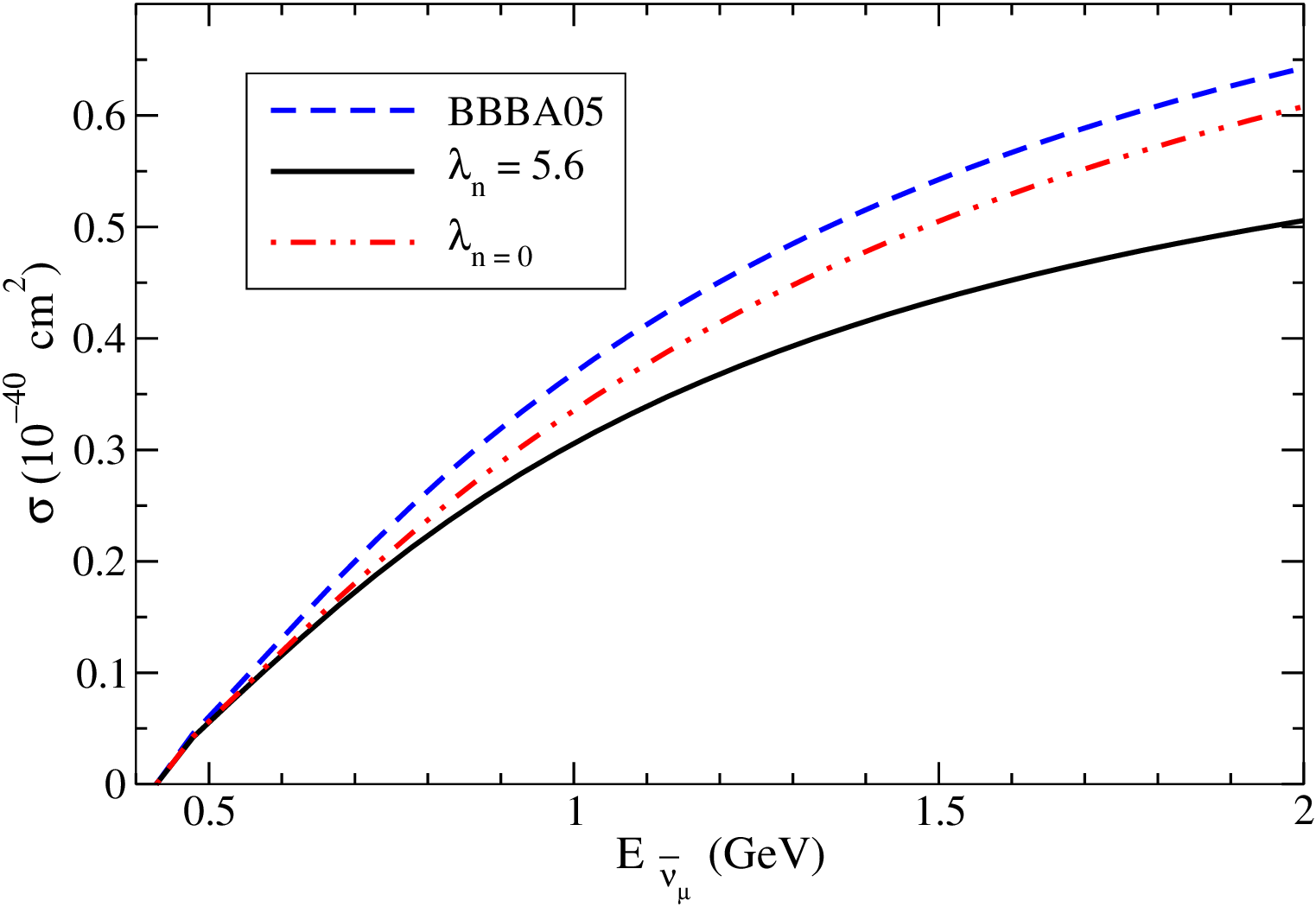}
\caption{(color online). $\sigma$ vs $E_{\bar\nu_\mu}$, for $\bar\nu_\mu + p \rightarrow \mu^+ + \Sigma^0$ process.
The results are presented with the 
  neutron form factors using Eq.\ref{eq:xin_def} with $\lambda_n$=0 and $\lambda_n$=5.6, as well as 
  the form factor given by Bradford et al.~\cite{Bradford:2006yz}.
}
\label{fg:xsec_p_mu_Sig_ff}     
\end{figure}

\begin{figure}\centering
\includegraphics[height=0.26\textheight,width=0.7\textwidth]{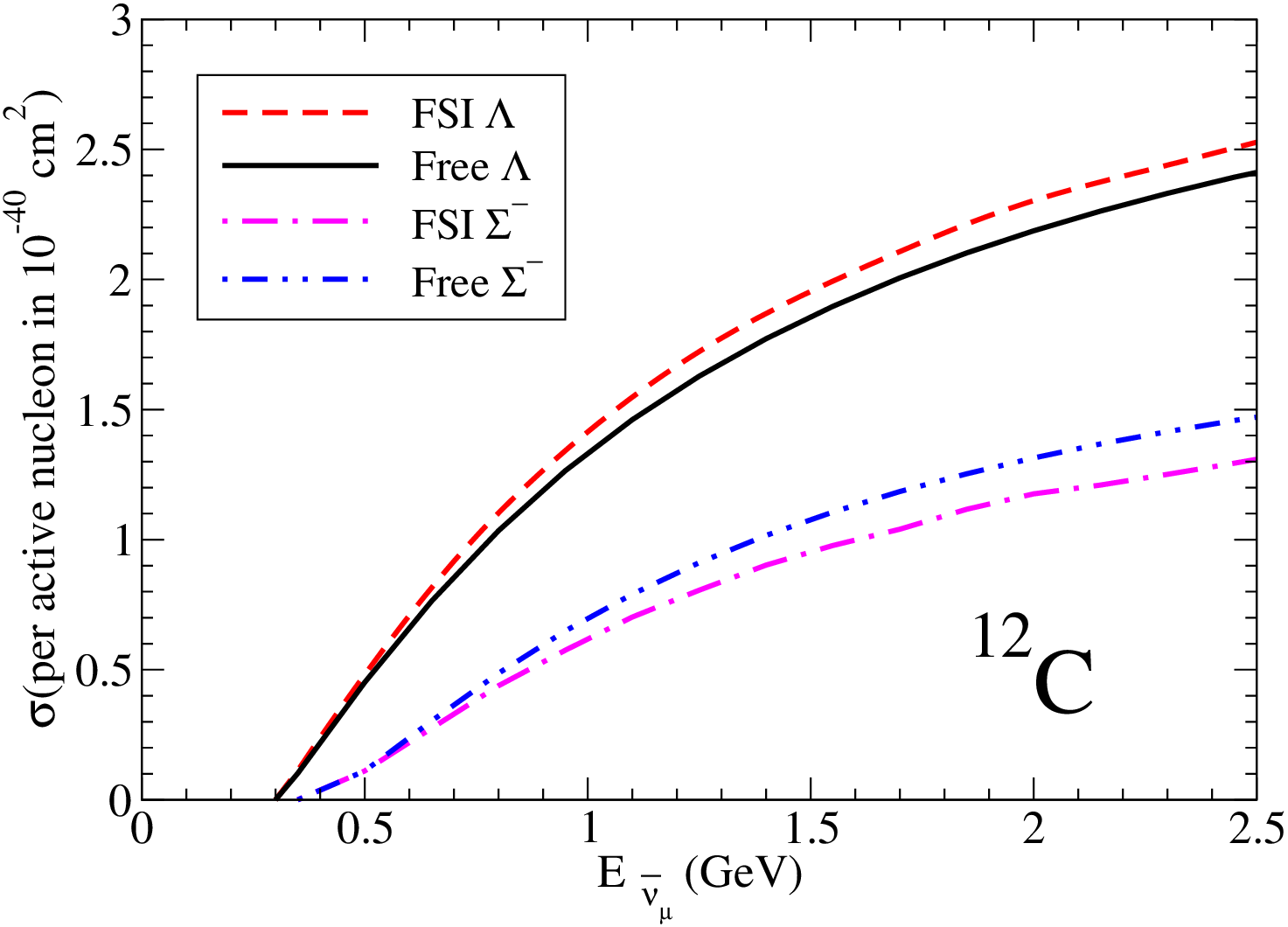}
\caption{(color online). $\sigma$(per active nucleon) vs $E_{\bar\nu_\mu}$ in $^{12}$C, for quasielastic hyperon production, 
for $\Lambda$ production(per proton) the results are shown without(solid) and with FSI(dashed line). Corresponding results 
 are shown for $\Sigma^-$ production(per neutron) without(dashed-double dotted line) and with FSI(dashed-dotted line).}
\label{fg:xsec_c12}
\end{figure}

\begin{figure}\centering
\includegraphics[height=0.26\textheight,width=0.7\textwidth]{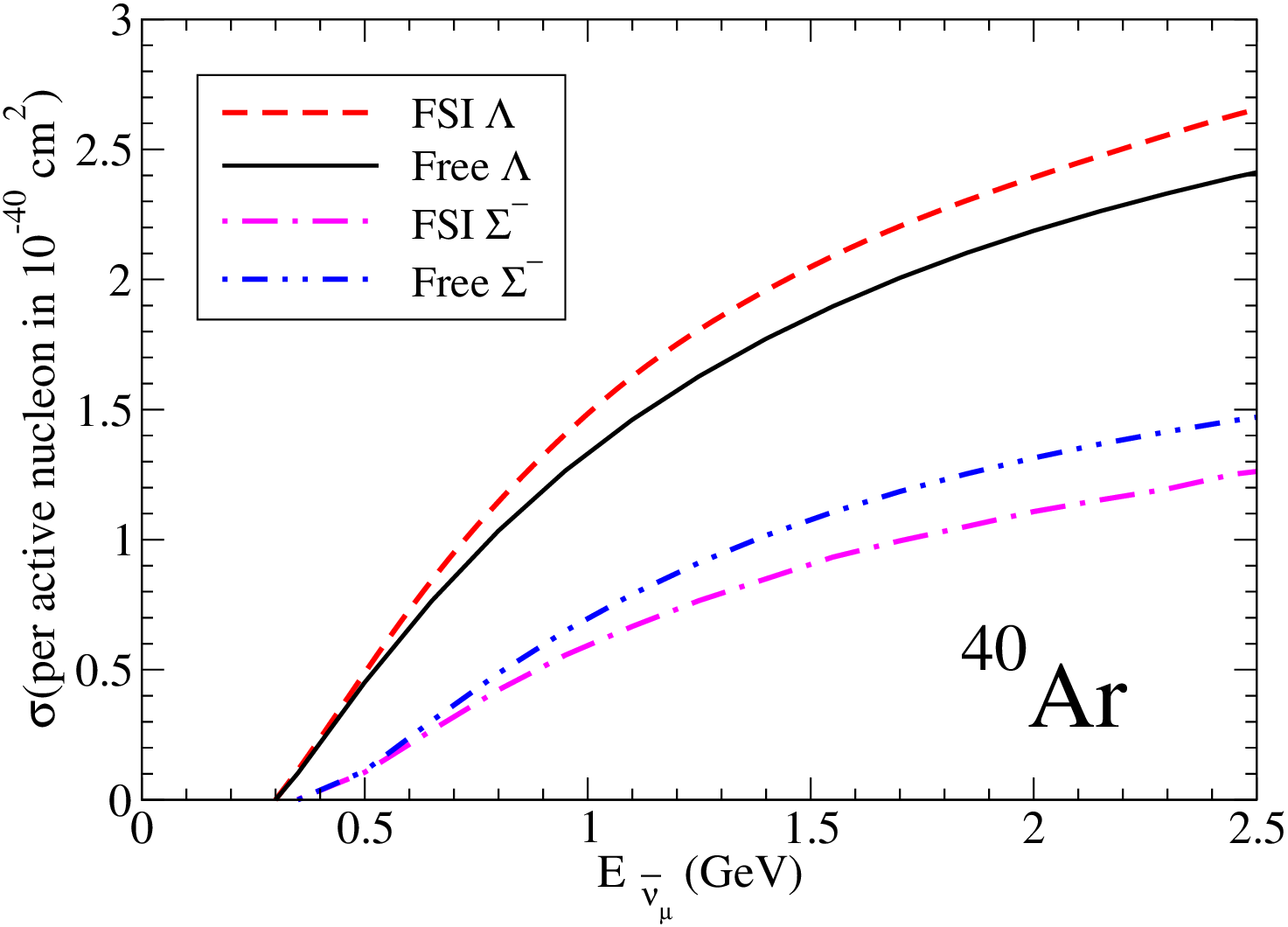}
\caption{(color online). $\sigma$(per active nucleon) vs $E_{\bar\nu_\mu}$ in $^{40}$Ar, for quasielastic hyperon production. 
Curves here have the same meaning as in Fig.\ref{fg:xsec_c12}.}
\label{fg:xsec_ar}
\end{figure}

\begin{figure}\centering
\includegraphics[height=0.26\textheight,width=0.7\textwidth]{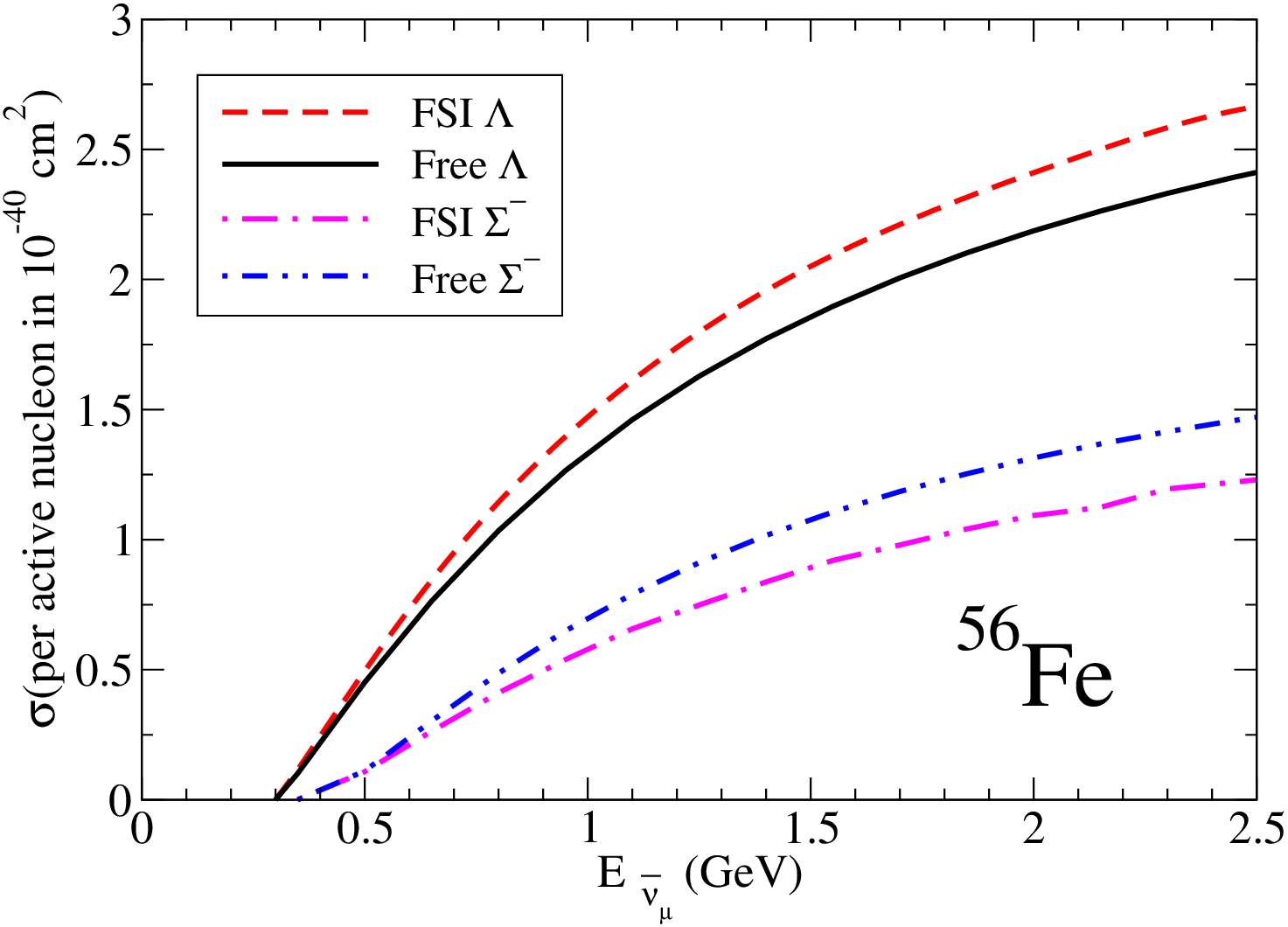}
\caption{(color online). $\sigma$(per active nucleon) vs $E_{\bar\nu_\mu}$ in $^{56}$Fe, for quasielastic hyperon production. 
Curves here have the same meaning as in Fig.\ref{fg:xsec_c12}.}
\label{fg:xsec_fe}
\end{figure}

\begin{figure}\centering
\includegraphics[height=0.26\textheight,width=0.7\textwidth]{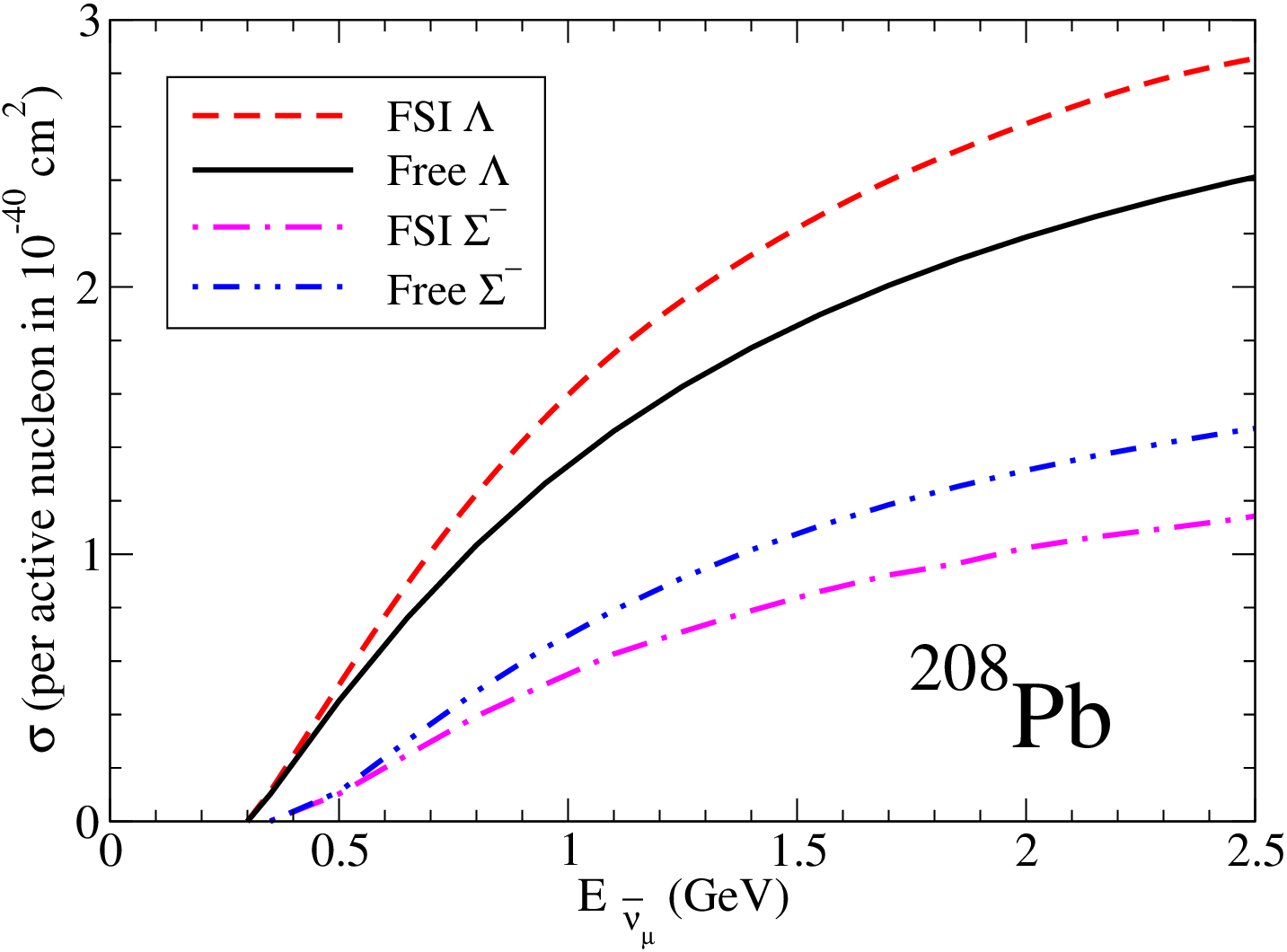}
\caption{(color online). $\sigma$(per active nucleon) vs $E_{\bar\nu_\mu}$ in $^{208}$Pb, for quasielastic hyperon production. 
Curves have the same meaning as in Fig.\ref{fg:xsec_c12}.}
\label{fg:xsec_pb}
\end{figure}

\begin{figure}\centering
\includegraphics[height=0.26\textheight,width=0.7\textwidth]{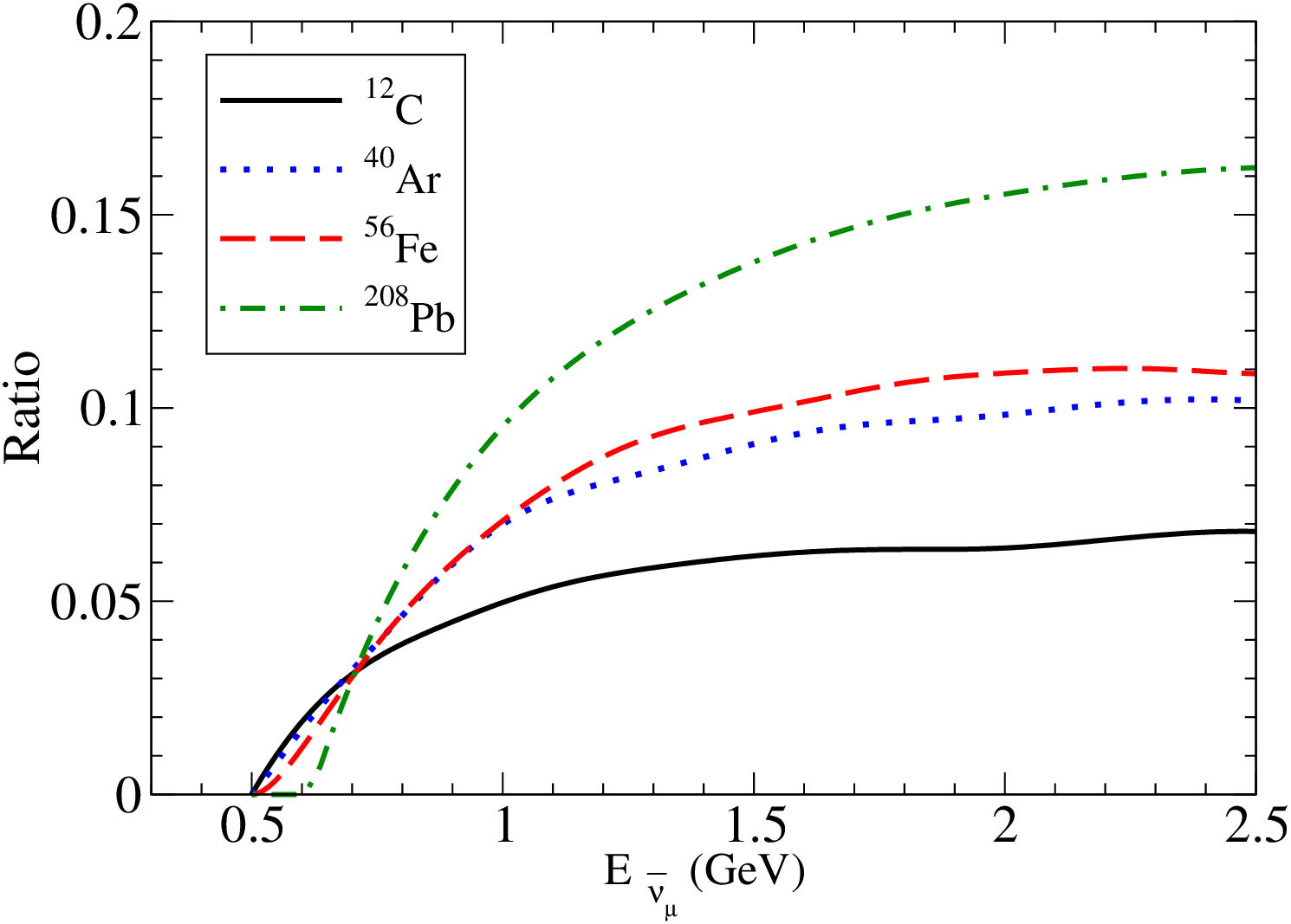}
\caption{(color online). Ratio of $\frac{\sigma(\Sigma^0) - \frac12 \sigma(\Sigma^-)}{\sigma(\Sigma^0)}$  for $\bar\nu_\mu$ induced interactions 
per active nucleon in nuclei with Final State Interaction(FSI) effect.
 Solid line is the result in $^{12}C$, dotted is the result in $^{40}Ar$, dashed line is the result in $^{56}Fe$ and dashed-dotted line is the result in $^{208}Pb$.}
\label{fg:pl}
\end{figure}

\begin{figure}\centering
\includegraphics[height=0.26\textheight,width=0.7\textwidth]{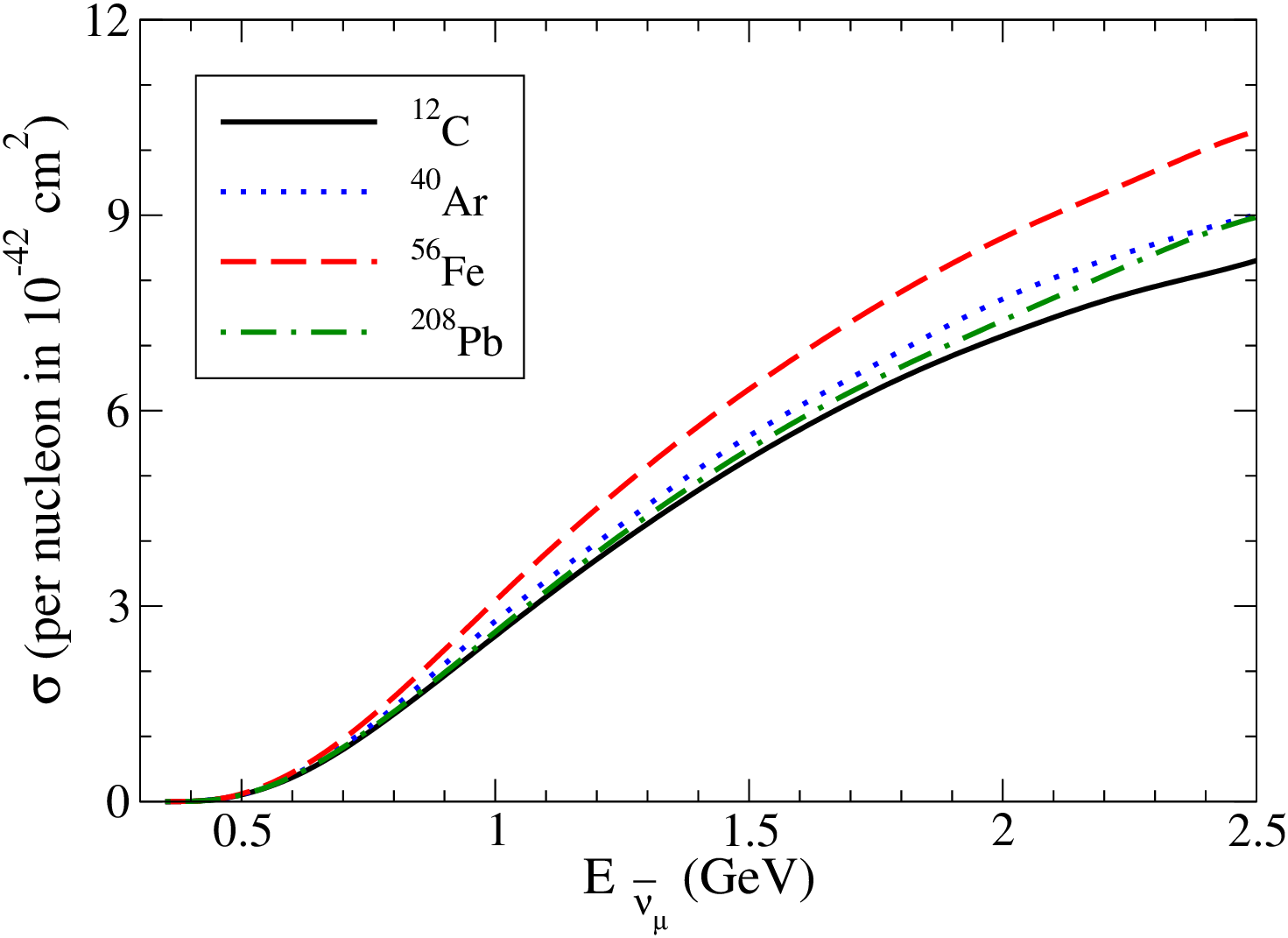}
\caption{(color online). $\sigma$ vs $E_{\bar\nu_\mu}$, for $\Sigma^+$ production(per nucleon) arising due to final state interaction effect of $\Lambda$ and $\Sigma^0$ hyperons in nuclei. 
Solid line is the result in $^{12}C$, dotted is the result in $^{40}Ar$, dashed line is the result in $^{56}Fe$ and dashed-dotted line is the result in $^{208}Pb$. }
\label{fg:xsec_sig_plus}
\end{figure}

\begin{figure}\centering
\includegraphics[height=0.26\textheight,width=0.7\textwidth]{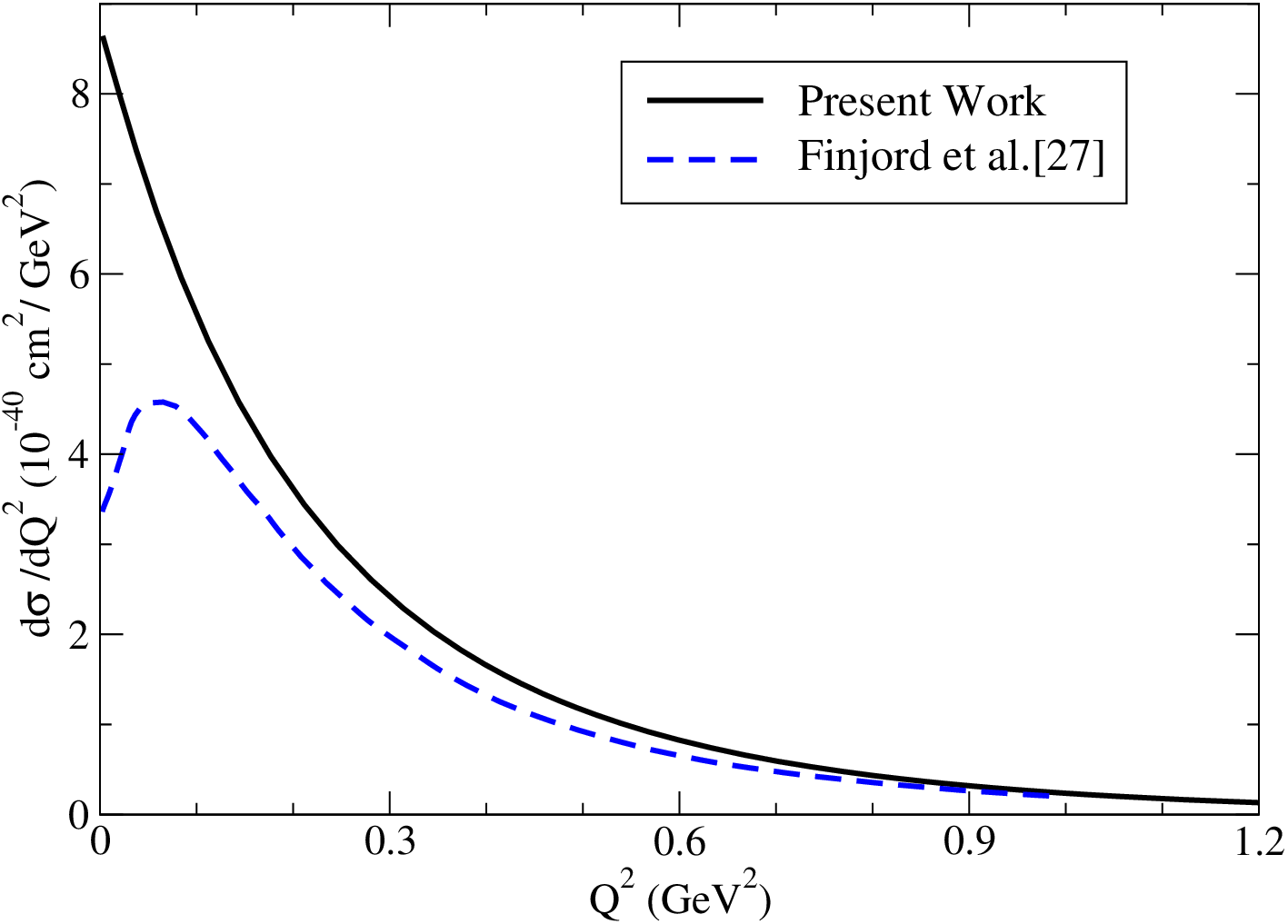}
\caption{(color online). $\frac{d\sigma}{dQ^2}$ vs $Q^2$ for $\bar\nu_\mu + p \rightarrow \mu^+ + \Lambda$ process at
$E_{\bar\nu_\mu}$=2GeV. Solid line is the result with the present model and the dashed line 
is the result of Finjord and Ravndal~\cite{Finjord:1975zy}.}
\label{fg:rav_lam}
\end{figure}

\begin{figure}\centering
\includegraphics[height=0.26\textheight,width=0.7\textwidth]{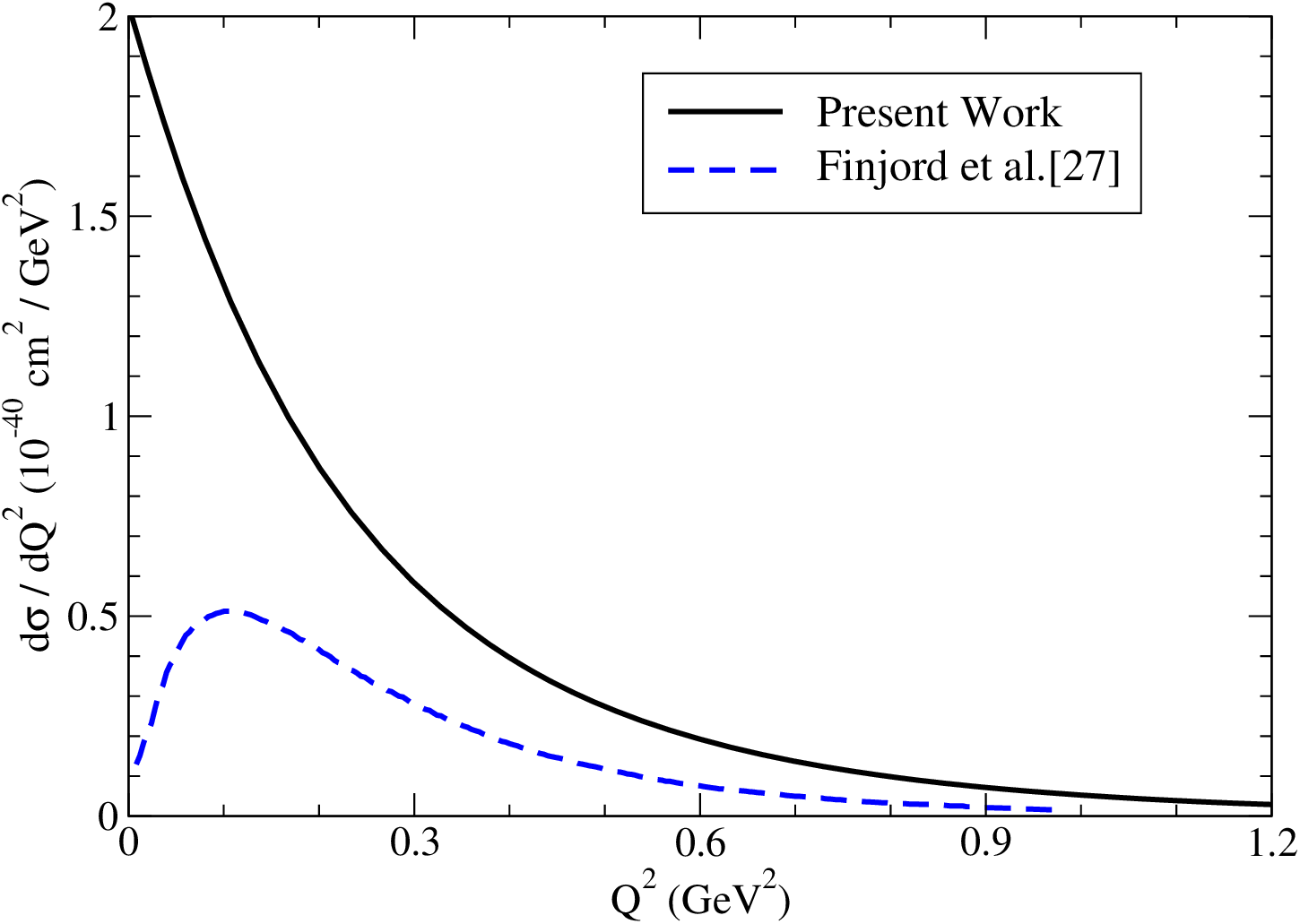}
\caption{(color online). $\frac{d\sigma}{dQ^2}$ vs $Q^2$ for $\bar\nu_\mu + p \rightarrow \mu^+ + \Sigma^0$ process 
at $E_{\bar\nu_\mu}$=2GeV. Solid line is the result with the present model and the dashed line 
is the result of Finjord and Ravndal~\cite{Finjord:1975zy}.}
\label{fg:rav_sig0}
\end{figure}

\begin{figure}\centering
\includegraphics[height=0.26\textheight,width=0.7\textwidth]{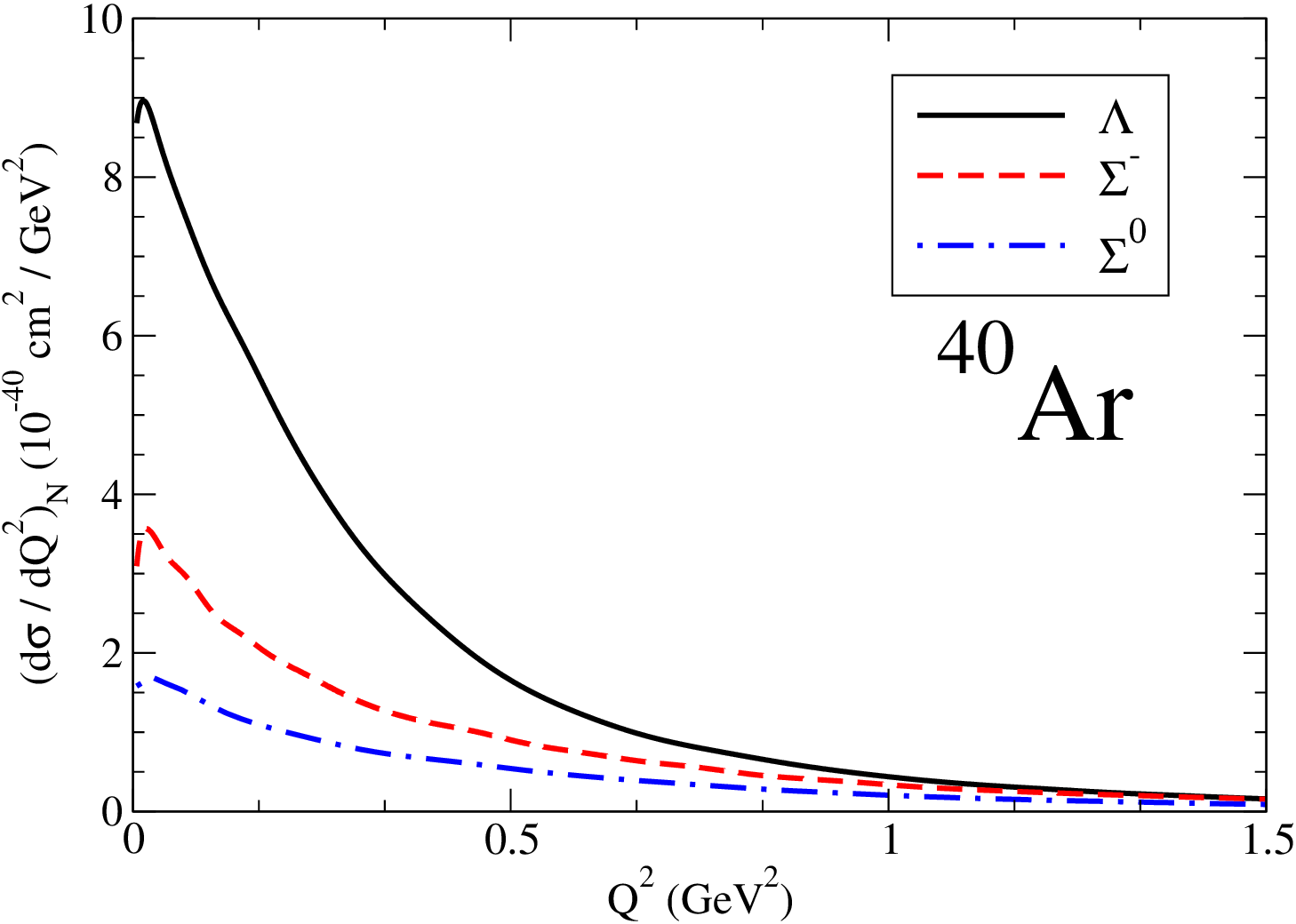}
\caption{(color online). $Q^2$ distribution(per active nucleon) for $\bar\nu_\mu + N \rightarrow \mu^+ + Y$ processes
(N=proton for $Y=\Lambda$ / $\Sigma^0$ and N=neutron
for $Y=\Sigma^-$ production) at  $<E_{\bar\nu_\mu}>$=3.6GeV corresponding to the average energy of ArgoNeuT experiment~\cite{Anderson:2011ce, Acciarri:2014isz}.}
\label{fg:q2_argoneut}
\end{figure}

\begin{figure}\centering
\includegraphics[height=0.26\textheight,width=0.4\textwidth]{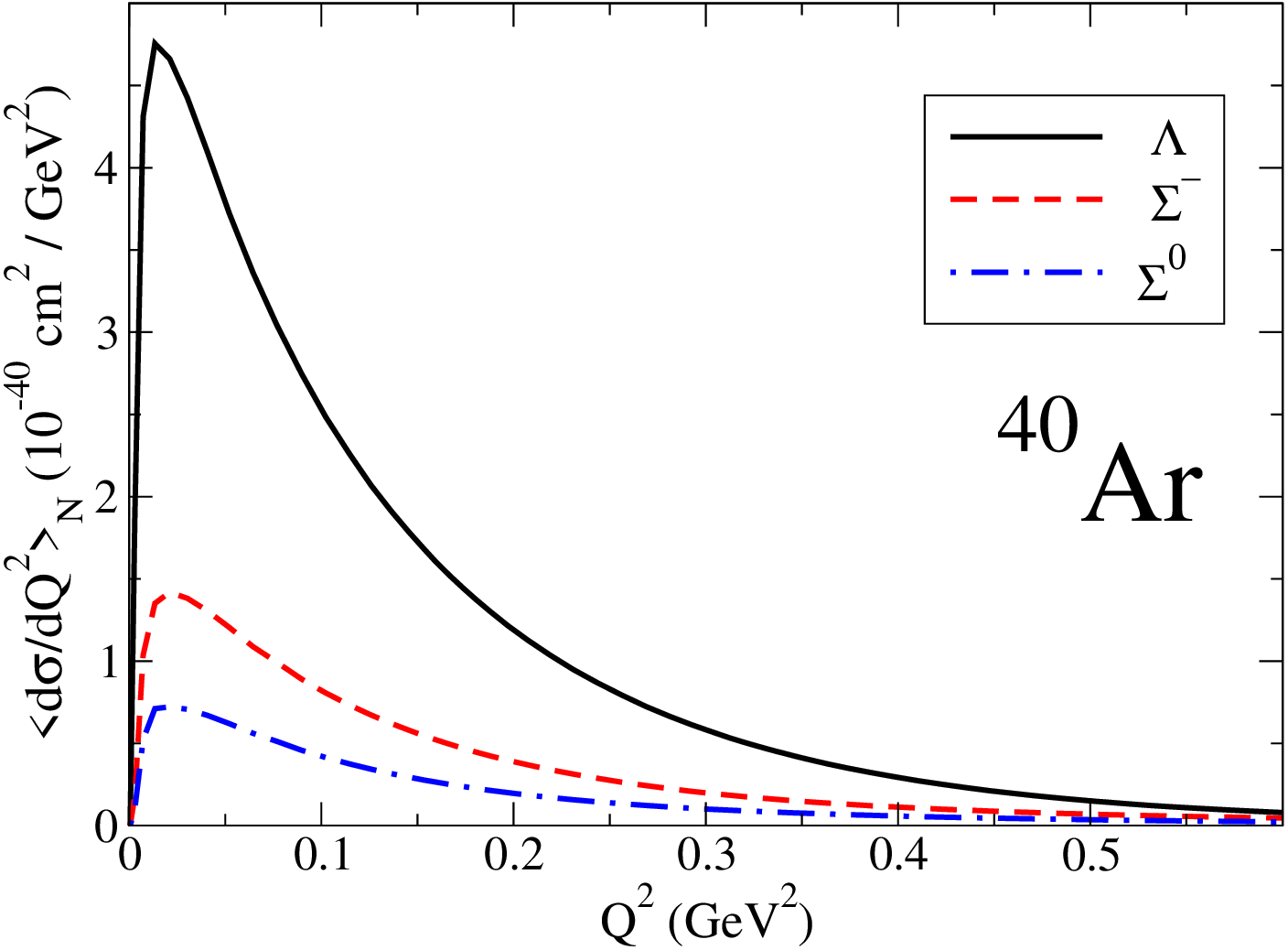}
\includegraphics[height=0.26\textheight,width=0.4\textwidth]{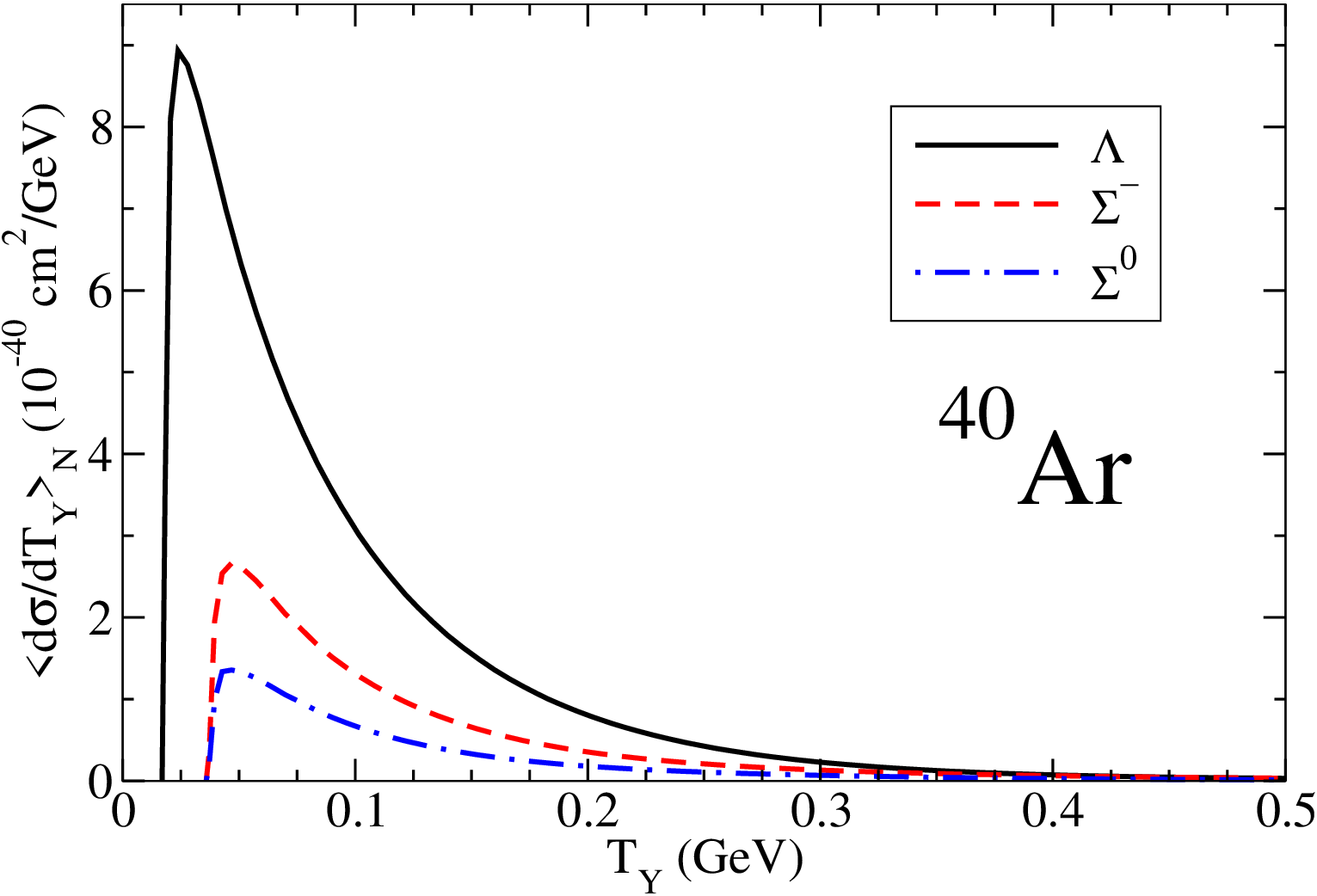}
\caption{(color online). $ \left \langle \frac{d\sigma}{dQ^2} \right \rangle_N $ vs $Q^2$ and $\left \langle \frac{d \sigma}{ d T_Y}\right \rangle_N$ vs
$T_Y$ in $^{40}Ar$ (per active nucleon N; N=proton for $Y=\Lambda$ / $\Sigma^0$ and N=neutron
for $Y=\Sigma^-$ production)  obtained by averaging $Q^2$-distribution and the kinetic energy distribution 
over the MicroBooNE~\cite{Ignarra:2011yq,Karagiorgi:2012jz} flux for the reactions given 
in Eq.~\ref{hyp-rec}. The results are presented with nuclear 
medium and final state interaction effects.}
\label{fg:q2ty_mb}
\end{figure}

\begin{figure}\centering
\includegraphics[height=0.26\textheight,width=0.3\textwidth]{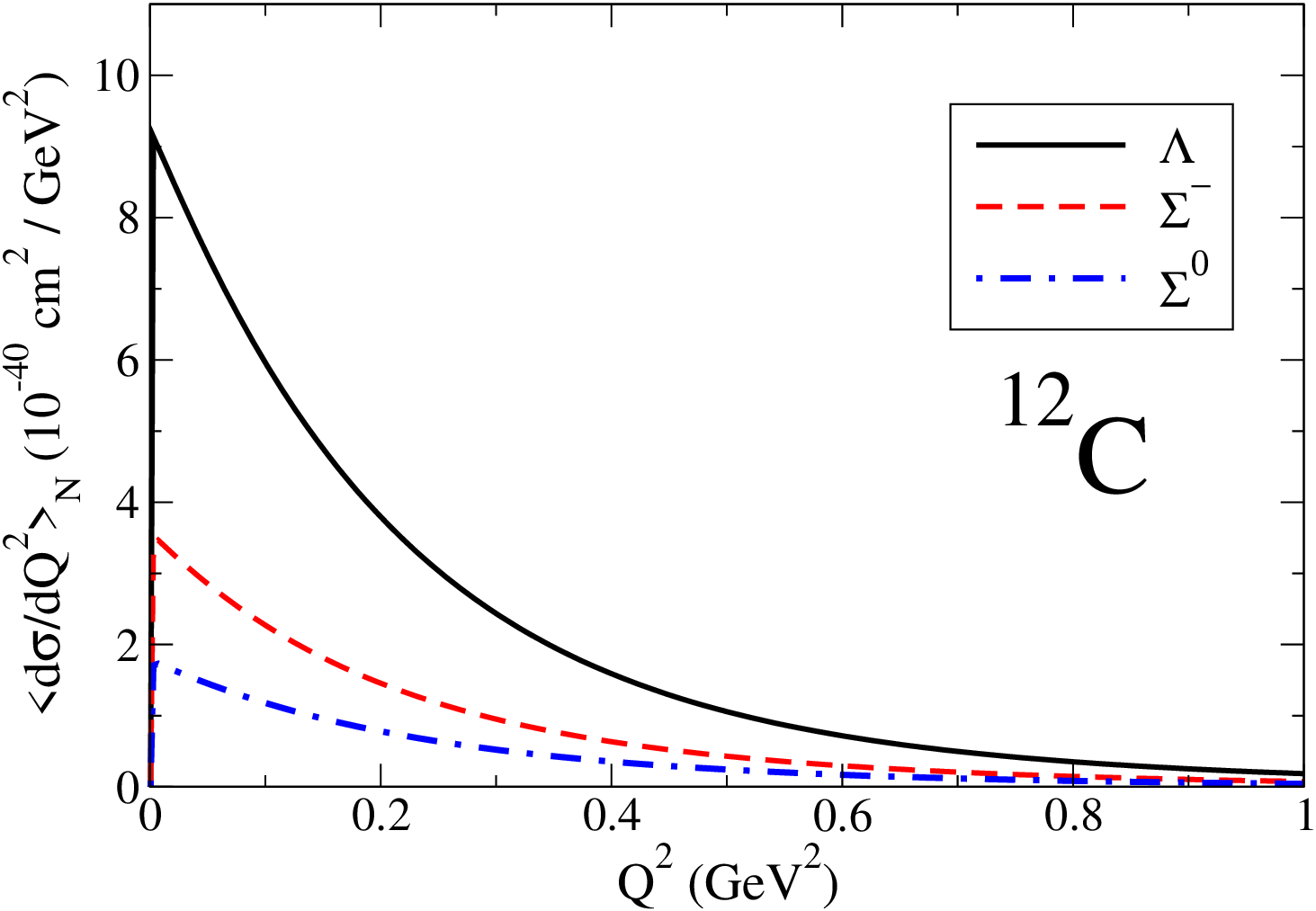}
\includegraphics[height=0.26\textheight,width=0.3\textwidth]{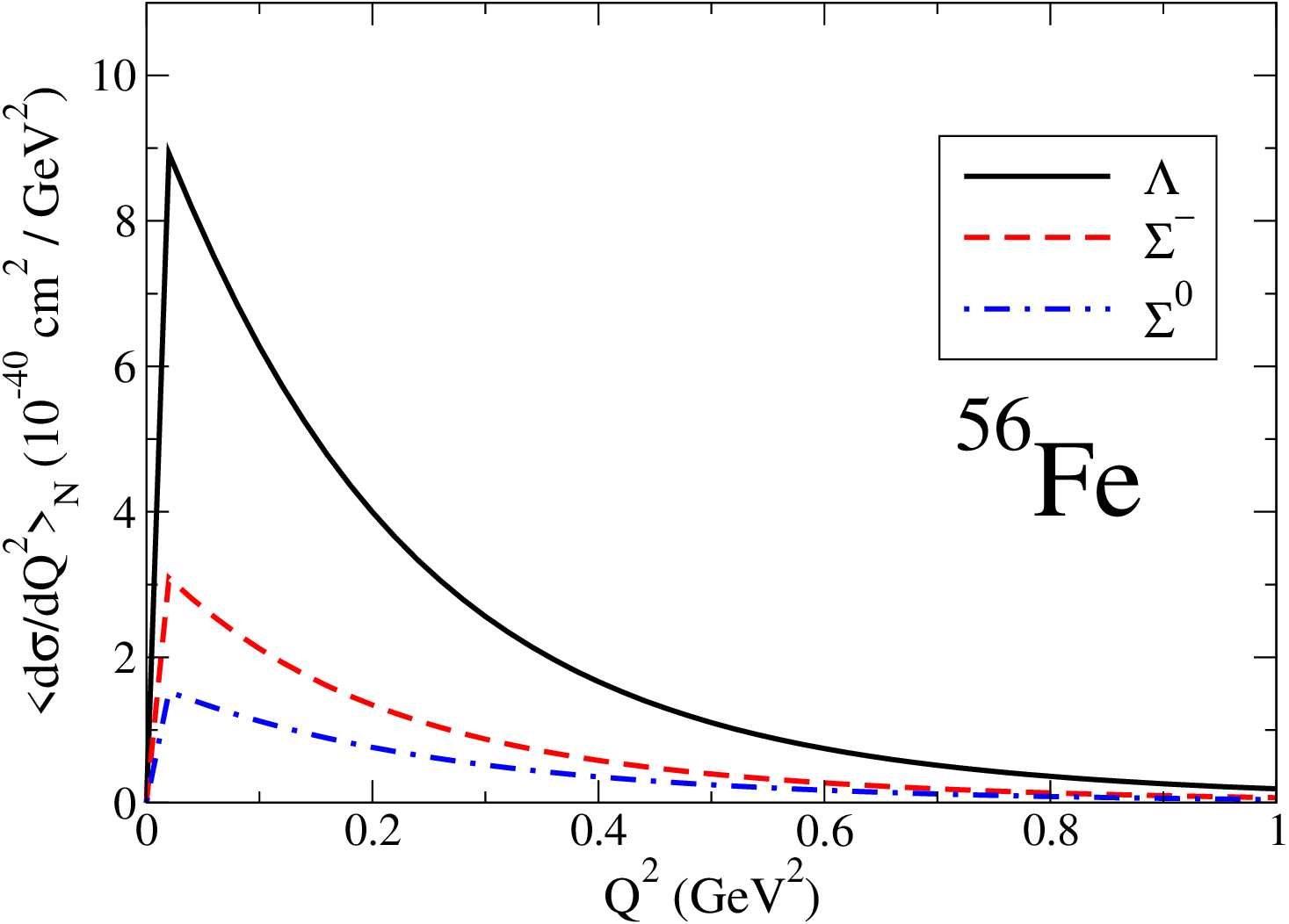}
\includegraphics[height=0.26\textheight,width=0.3\textwidth]{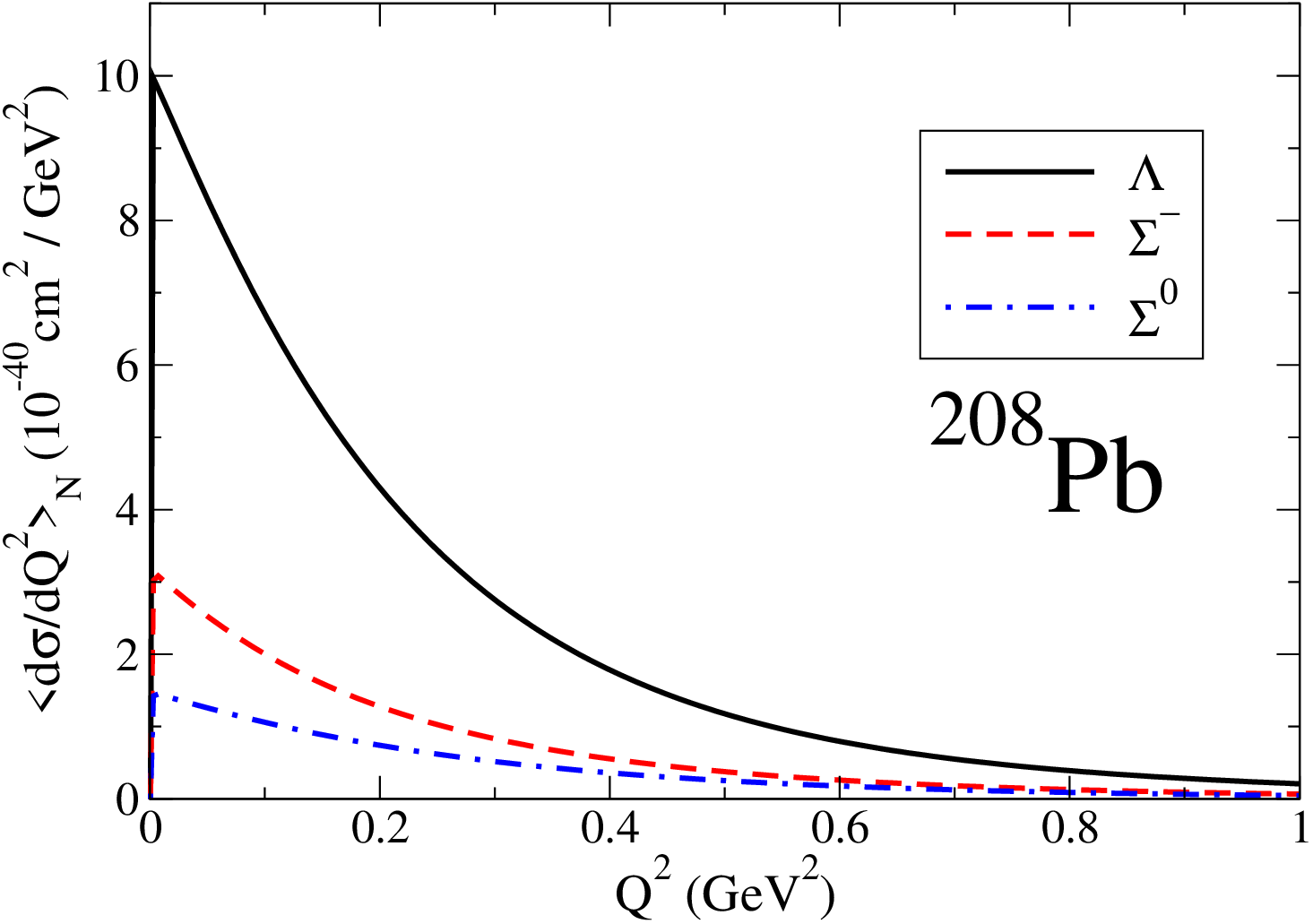}
\caption{(color online). $ \left \langle \frac{d\sigma}{dQ^2} \right \rangle_N $ vs $Q^2$ in $^{12}C$, $^{56}Fe$ and $^{208}Pb$ nuclear
targets(per active nucleon) obtained by averaging $Q^2$-distribution over the MINER$\nu$A~\cite{Fields:2013zhk} 
flux for the reactions given in Eq.~\ref{hyp-rec}. The results are presented with nuclear 
medium and final state interaction effects.}
\label{fg:q2_minerva}
\end{figure}

\section{Results and Discussion}\label{sec:results}
The numerical results for the quasielastic production of $\Sigma^0,\; \Sigma^-$ and $\Lambda$ 
hyperons induced by antineutrinos from the nucleon targets have been presented using Eqs.~\ref{crosv.eq} and \ref{matrix1} 
with the vector and axial vector form factors given by Eqs.\ref{Vec_FF} and \ref{Axi_FF}.
The results have been then applied to calculate the hyperon production 
in nuclear targets like $^{12}C$, $^{40}Ar$, $^{56}Fe$ and $^{208}Pb$ where nuclear medium effects and final
state interaction effect due to hyperon-nucleon interaction in the nuclear medium have been considered. 
The numerical results for the $Q^2$ distribution have also been presented. 
In the case of free nucleon the results for $\bar \nu_\mu + p \rightarrow \mu^+ + \Lambda $ and 
$\bar \nu_\mu + p \rightarrow \mu^+ + \Sigma^0 $ have been presented. The cross sections for 
$\bar \nu_\mu + n \rightarrow \mu^+ + \Sigma^- $ are related to $\bar \nu_\mu + p \rightarrow \mu^+ + \Sigma^0 $
by a simple relation and is given by $\sigma(\bar \nu_\mu p \rightarrow \mu^+ \Sigma^0) =
\frac12 \sigma(\bar \nu_\mu n \rightarrow \mu^+ \Sigma^-)$, 
which is modified due to final state interaction in the medium 
and has been discussed in some detail. Similarly, $\Sigma^+$ is not produced from a free nucleon but can be produced 
through the final state interactions in the nuclear medium for which results have been presented. 
\subsection{Total Cross Section}\label{sec:sigma}
In Fig.~\ref{fg:xsec_comp}, we have presented the results for ${\bar\nu}_\mu$ induced
$\Lambda$ production from free proton in the energy region of $E_{{\bar\nu}_\mu} < 10$GeV and compared them with the
quark model calculations of Wu et al.\cite{Wu:2013kla} and Finjord and Ravndal~\cite{Finjord:1975zy} as well as 
with the  experimental results from Gargamelle bubble chamber at CERN~\cite{Eichten:1972bb,Erriquez:1978pg,Erriquez:1977tr} 
using Propane with a small admixture of Freon and from Serpukhov SKAT Bubble Chambers~\cite{Brunner:1989kw} using Freon 
and results from BNL experiment using Hydrogen target~\cite{Fanourakis:1980si}. The results of theoretical calculations performed by Erriquez et al.~\cite{Erriquez:1978pg}, Brunner et
al.\cite{Brunner:1989kw} and Kuzmin and Naumov~\cite{Kuzmin:2008zz} are based on the prediction from Cabibbo theory. 

In Fig.~\ref{fg:xsec_comp}, we have used the currently favored value of $M_A=1.03$GeV. 
On the other hand Erriquez {\it et al.}~\cite{Erriquez:1978pg} have used $M_A = 0.883$GeV and 
Brunner {\it et al.}~\cite{Brunner:1989kw} have used $M_A = 1.0$GeV and 
obtained slightly smaller values of the total cross sections in better agreement with the experimental results. 
In a recent calculation Kuzmin and Naumov~\cite{Kuzmin:2008zz} have used $M_A = 0.999$GeV with the BBBA~\cite{Bodek:2007ym}  
parameterization of the vector form factors to obtain a good agreement with the experimental results. 
While determining the vector form factors in connection
with $|\Delta S|= 1$ reaction as well as semi leptonic decay of hyperons, 
it is worth mentioning the suggestion of Gaillard and Sauvage~\cite{Gaillard:1984ny}
that a rescaled value of $M_V = 0.97$ GeV instead of 
$M_V = 0.84$ GeV should be used while implementing the SU(3) symmetry.
A larger value of $M_V $ will make the agreement with the experimental results worse. 
In Fig.~\ref{fg:xsec_comp}, we have also shown the theoretical results obtained in the quark models of
Finjord and Ravndal~\cite{Finjord:1975zy} and Wu et al.~\cite{Wu:2013kla}. 
While Finjord and Ravndal~\cite{Finjord:1975zy} use a covariant relativistic quark model, Wu et al.~\cite{Wu:2013kla} 
use a non-relativistic quark model. In general quark model underestimates the 
total cross section for $\bar \nu_\mu p \rightarrow \mu^+ \Lambda$  as compared to Cabibbo theory and/or experimental results. 
The results of Wu et al.~\cite{Wu:2013kla} are too small. In fact the results of Wu et al.\cite{Wu:2013kla}
are very sensitive to the variation in the harmonic oscillator parameters of the non-relativistic quark model 
as discussed by them. In view of this a quark model analysis of hyperon production in electroweak processes is highly desirable.

In the numerical calculations, using Cabibbo theory, the magnetic form factor $f_2 (q^2)$ is determined using SU(3) symmetry and is given by (see Table-\ref{tab2})
\begin{eqnarray}\label{eq:f3f2_lam}
 f_2^{p \rightarrow \Lambda} (q^2) = - \sqrt \frac{3}{2} f_2^p (q^2)
\end{eqnarray}
However, instead of Eq.~\ref{eq:f3f2_lam}, if we choose to implement the 
SU(3) symmetry at the level of $\frac{f_2 (q^2) }{ M_Y}$ as done by Cabibbo {\it et al.}~\cite{Cabibbo:2003cu},
we would obtain with our definitions of matrix element defined in Eq.~\ref{matrix1} 
\begin{eqnarray}
\frac{f_2^{p \rightarrow \Lambda} (q^2) }{ M_\Lambda + M} = - \sqrt \frac{3}{2} \frac{f_2^p (q^2) }{M_n + M_p }
\end{eqnarray}
leading to 
\begin{eqnarray}\label{eq:f3f2_lam_modified}
f_2^{p \rightarrow \Lambda} (q^2)  = - \sqrt \frac{3}{2} \frac{M_\Lambda + M}{ M_n + M_p } f_2^p (q^2)  
\end{eqnarray}
and similarly
\begin{eqnarray}\label{eq:f3f2_sig_modified}
f_2^{p \rightarrow \Sigma^0} (q^2)  =  \frac{-1}{\sqrt2} \frac{M_\Sigma + M}{ M_n + M_p }  (f_2^p(q^2) + 2 f_2^n(q^2)).
\end{eqnarray}

This implies a 10-15 $\%$ variation in the values of the form factor $f_2(q^2)$. We find that the total cross sections
are not affected by this variation in the $f_2(q^2)$  form factor.
 This is consistent with the results of Dworkin et al.\cite{Dworkin:1990dd} who have 
 analyzed the semileptonic decay of $\Lambda$ hyperons using SU(3) symmetry and took two different values
for weak-magnetism coupling $\omega(=\frac{f_{2}(0)}{f_{1}(0)})$ viz. $\omega = 0.15$ 
and $\omega = 0.97$, and find that the decay rates are not affected. Note that their definition of the 
transition matrix element involving the $f_2(q^2)$ term is slightly different from ours.

In Fig.~\ref{fg:xsec_p_mu_Sig}, we present our results for the reaction  $\bar \nu_\mu + p \rightarrow \mu^+ + \Sigma^0 $ 
and compare them with the results of Wu et al.~\cite{Wu:2013kla} and Finjord and Ravndal~\cite{Finjord:1975zy}.  
We have also shown the experimental result 
 obtained in the bubble chamber Gargamelle experiment by Erriquez et al.~\cite{Erriquez:1978pg}.
 While the quark model results of Finjord and Ravndal~\cite{Finjord:1975zy}
give reasonable description of the observed data, the results of Wu et al.~\cite{Wu:2013kla} (dotted line) 
underestimates them and the Cabibbo theory (solid line) overestimates them. It is worth noting that 
the cross sections for $\bar \nu_\mu p \rightarrow \mu^+ \Sigma^0 $ and $\bar \nu_\mu n \rightarrow \mu^+ \Sigma^- $ 
depend upon the neutron form factors and is found to be quite sensitive to the value used for this
form factor. For example, using $\xi=1$ in Eq.\ref{eq:xin_def} yields a cross section of $0.65 \times 10^{-40} \;cm^2$ as compared 
to the cross section of $0.5 \times 10^{-40} \;cm^2$ at $E_{\bar\nu_\mu}=2GeV$ for $\bar \nu_\mu p \rightarrow \mu^+ \Sigma^0 $. 
It may be seen that the cross sections for $\bar \nu_\mu p \rightarrow \mu^+ \Sigma^0 $ and $\bar \nu_\mu n \rightarrow \mu^+ \Sigma^- $ reactions are 
quite sensitive to the choice of the neutron form factor $G_E^n(q^2)$. The dependence of the various parameterization of the neutron form factor $G_E^n(q^2)$ on the cross section may be seen 
from Fig.~\ref{fg:xsec_p_mu_Sig_ff}, where we have presented the results of $\sigma$ vs $E_{\bar\nu_\mu}$, for $\bar\nu_\mu + p \rightarrow \mu^+ + \Sigma^0$ reaction. These
results are presented with the neutron form factors using Eq.\ref{eq:xin_def} with $\lambda_n$=0 and $\lambda_n$=5.6, as well as the form factor given by Bradford et al.~\cite{Bradford:2006yz}.
 Similar is the dependence for $\bar \nu_\mu n \rightarrow \mu^+ \Sigma^- $  reaction.

We have also obtained  the results for the total cross section $\sigma(E_{\bar \nu_\mu})$ vs $E_{\bar \nu_\mu}$ 
for various nuclei of interest like
$^{12}C$, $^{40}Ar$, $^{56}Fe$ and $^{208}Pb$ relevant to ongoing (anti)neutrino experiments at 
T2K, MicroBooNE and MINER$\nu$A. The results for $\Lambda$ and $\Sigma^-$ production 
are  shown for $^{12}C$, $^{40}Ar$, $^{56}Fe$ and $^{208}Pb$ in Figs.~\ref{fg:xsec_c12}-\ref{fg:xsec_pb}. 
We find that nuclear medium effects due to Pauli blocking are very small. 
However, the final state interactions due to $\Sigma-N$ and $\Lambda - N$ interactions in various channels 
tend to increase the $\Lambda$ production and decrease the $\Sigma^-$ production. 
 The quantitative increase(decrease) in $\Lambda(\Sigma)$ yield due to FSI increases with the increase in nucleon 
number. The $\Sigma^-$ and $\Sigma^0$
production are separately affected and the relation 
$ \sigma \left(  \bar \nu_\mu + p \rightarrow \mu^+ + \Sigma^0   \right) = 
\frac12  \sigma \left(  \bar \nu_\mu + n \rightarrow \mu^+ + \Sigma^-   \right) $
is modified in the nucleus due to the presence of other nucleons. 

In Fig.~\ref{fg:pl}, we show $ {\cal R } =\frac{\sigma(\Sigma^0) - \frac12 \sigma(\Sigma^-) }{\sigma(\Sigma^0)}$, 
for the cross sections obtained per active nucleon 
as a function of antineutrino energy in $^{12}C$, $^{40}Ar$, $^{56}Fe$ and $^{208}Pb$. We see that at  low energies the value of ${\cal R }$ is higher
for isoscalar nuclei like $^{12}C$ than for non-isoscalar nuclei
 like $^{56}Fe$, $^{208}Pb$, etc. At higher energies the value of ${\cal R }$
 increases with the nucleon number. At low energies the $\Lambda N \rightarrow \Sigma N$ 
transitions are kinematically inhibited and the relative changes in $\Sigma^-$ and $\Sigma^0$ productions due to 
final state interaction effects are governed by inelastic $\Sigma N \rightarrow \Lambda N$ and charge exchange $\Sigma N \rightarrow \Sigma N$ reactions.

For charge exchange reactions, $\Sigma^0 N$  cross sections are smaller than $\Sigma^- N$ 
cross sections\footnote{See Appendix of Ref.~\cite{Singh:2006xp}.} and the initial weak production of
$\Sigma^0$ is half of $\Sigma^-$ production. 
In view of this the relative yields of $\Sigma^-$ and $\Sigma^0$ due to FSI are dominantly determined by
$\Sigma^- p \rightarrow \Lambda n$ and $\Sigma^- p\rightarrow \Sigma^0 n$ reactions. 
These reactions deplete more $\Sigma^-$ in the case of isoscalar nucleus as compared to non-isoscalar nucleus 
due to higher Fermi energy of neutrons than protons, making ${\cal R }$ larger for $^{12}C$ than $^{208}Pb$. 
At higher energies where all the inelastic processes like $\Lambda N \rightarrow \Sigma N$ and $\Sigma N \rightarrow \Lambda N $
and charge exchange reactions like $\Sigma N \rightarrow \Sigma^\prime N^\prime $ contribute, the value of ${\cal R }$ 
increases with the nucleon number mainly
due to larger initial production of $\Sigma^-$ and larger $\Sigma^- p \rightarrow \Lambda n$, 
$\Sigma^- n \rightarrow \Sigma^0 p$ cross sections. These effects are quantitatively seen in Fig.~\ref{fg:pl}.

 We also see the appearance of
$\Sigma^+$ due to final state interaction processes like 
$\Lambda p \rightarrow \Sigma^+ n$ and $\Sigma^0 p \rightarrow \Sigma^+ n  $. In Fig.~\ref{fg:xsec_sig_plus}, we present the results for 
the cross section for $\Sigma^+$ production as a function of antineutrino energy in various nuclei. 
 The cross section per nucleon (i.e. $ \sigma/ \frac{N+Z}{2} $) for $\Sigma^+$ production increases with the increase in 
proton number except for $^{208}Pb$ where we see a suppression as compared to $^{56}Fe$. 
This may be due to considerably higher Fermi energy of neutrons than protons in $^{208}Pb$ which inhibits the production of 
$\Sigma^+$ through $\Lambda p \rightarrow \Sigma^+ n$ and $\Sigma^0 p \rightarrow \Sigma^+ n  $
reactions in $^{208}Pb$ due to threshold considerations.
It will be interesting to test 
these productions whenever the experimental results are available in future.

In Figs.~\ref{fg:rav_lam}-\ref{fg:rav_sig0}, we show $ d \sigma / d Q^2$ for $\bar \nu_\mu + p \rightarrow \mu^+ + \Lambda$
and $\bar \nu_\mu + p \rightarrow \mu^+ + \Sigma^0  $ processes at $E_{\bar \nu} =2$ GeV and compare the results with the 
quark model calculation of Finjord and Ravndal~\cite{Finjord:1975zy}. We observe that the quark model gives smaller cross sections
 specially in the forward direction. 
 
In Fig.~\ref{fg:q2_argoneut}, we show the $Q^2$ distribution per nucleon  for $^{40}Ar$ 
at antineutrino energy of 3.6 GeV which is the  average energy of antineutrinos 
for ArgoNeuT~\cite{Anderson:2011ce, Acciarri:2014isz} experiment. 
We observe that the peak of $ d \sigma / d Q^2$ distribution shifts to lower $Q^2$ 
with the increase in the energy of the antineutrino beam.
While, in  Fig.~\ref{fg:q2ty_mb} we have presented the 
results for $Q^2$ distribution per active nucleon convoluted with the MicroBooNE~\cite{Ignarra:2011yq,Karagiorgi:2012jz} 
flux for $^{40}Ar$ nuclear target using 
  \begin{equation}\label{eq:avg_q2}
   \langle\frac{d \sigma}{ d Q^2}\rangle=\frac{\int_0^\infty \frac{d \sigma}{ d Q^2} 
   \phi(E_{\bar\nu_\mu}) dE_{\bar\nu_\mu}}{\int_0^\infty \phi(E_{\bar\nu_\mu}) dE_{\bar\nu_\mu}},
  \end{equation}
where $\phi(E_{\bar\nu_\mu})$ is $\bar\nu_\mu$ flux. 
In the right panel of Fig.~\ref{fg:q2ty_mb}, we have also presented the flux averaged kinetic energy 
distribution of hyperon i.e. $ \left \langle  \frac{d \sigma}{ d T_Y} \right \rangle  $ per nucleon  vs $T_Y$, 
using the MicroBooNE~~\cite{Ignarra:2011yq,Karagiorgi:2012jz} flux for $^{40}Ar$ nuclear target, where $T_Y$ is 
the kinetic energy of the outgoing hyperon. The kinetic energy distribution of hyperon is related to the 
$Q^2$ distribution by a simple relation between $Q^2$ and $T_Y$ (i.e. $Q^2=2 M (T_Y + M_Y) - M^2 - M_Y^2$).
Therefore, we have not shown the  kinetic energy distribution for the other cases 
which can be read from $Q^2$ distribution.

Similarly, in Fig.~\ref{fg:q2_minerva}, we have shown flux averaged $Q^2$ distribution i.e. 
$ \left \langle  \frac{d \sigma}{ d Q^2} \right \rangle  $ per active nucleon vs $Q^2$  
convoluted with  MINER$\nu$A~\cite{Fields:2013zhk} flux for $^{12}C$, $^{56}Fe$ and $^{208}Pb$ nuclear targets  using Eq.~\ref{eq:avg_q2}. 
We must point out that  because of the final state interaction effects 
there is an enhancement in the $Q^2$ distribution as well as in the kinetic energy distribution in
the $\Lambda$ production with nuclear medium and final state interaction effects than the results 
obtained for the free case(not shown), and decrease in the case of $\Sigma$ production. 
These results would also be useful in the analysis of atmospheric 
(anti)neutrino experiments like the proposed one at India Neutrino Observatory(INO)~\cite{ino} where $^{56}Fe$ is 
planned to be used as target.

 \section{Summary and Outlook}\label{sec:summary}
In this paper, we have studied quasielastic production of $\Sigma$ and $\Lambda$ hyperons from nucleons and 
nuclear targets induced by antineutrinos through $|\Delta S| =1$ weak charged currents. 
The calculations have been done using Cabibbo theory along with the symmetry properties of weak $|\Delta S| =1$ hadronic 
currents and various transition form factors have been determined using SU(3) symmetry, G-invariance and 
conserved vector current. Some SU(3) breaking effects are included through the use of physical masses for all 
the hadrons belonging to SU(3) octet. There is some indication of nuclear medium effects in the
work of Erriquez et al.~\cite{Erriquez:1978pg,Erriquez:1977tr}
where they quote the results for the  cross sections separately for
free and bound (including free) nucleons which shows a reduction due to the nuclear medium effects 
though consistent with no medium effects within the statistical errors.
On the other hand the experimental analysis of Eichten et al.~\cite{Eichten:1972bb} 
implies an increase of about 5$\%$ in lambda production. 
In view of these results and future experiments to be performed 
at low and medium energies it is important to study the nuclear medium effects in 
quasielastic production of hyperons.

The results for the total cross section, $Q^2$-distribution as well as the kinetic energy
distribution of hyperons would be relevant for the experiments 
like MINER$\nu$A~\cite{Fields:2013zhk}, MicroBooNE~\cite{Ignarra:2011yq,Karagiorgi:2012jz}, 
ArgoNeuT~\cite{Anderson:2011ce, Acciarri:2014isz}, 
INO~\cite{ino}, etc.
Therefore, we have presented the results for $\bar \nu_\mu + p \rightarrow \mu^+ + \Lambda $,
$\bar \nu_\mu + p \rightarrow \mu^+ + \Sigma^0 $ 
and $\bar \nu_\mu + n \rightarrow \mu^+ + \Sigma^- $ processes on nuclear targets like $^{12}C$, $^{40}Ar$, $^{56}Fe$ and $^{208}Pb$ 
which are being used in the 
present and proposed experiments. 
The theoretical results on free nucleons have been compared with predictions of quark model calculations where available. 
A comparison with old experimental results from Gargamelle, BNL and SKAT collaborations has been presented. 
 The nuclear medium effects due to Pauli blocking and final state interaction effect due to hyperon-nucleon 
interactions in the presence of other nucleons in the nuclear medium have been included. 
The deviation from SU(3) symmetric predictions for $ \Sigma^-$ and $ \Sigma^0$ production and appearance 
of $ \Sigma^+$ production due to final state interactions have been studied using Monte Carlo 
simulation of final state interactions using experimental hyperon-nucleon scattering cross sections. 
While the nuclear medium effects due to Pauli blocking are small, the final state interactions lead to an increase 
of $\Lambda$ production and decrease of $ \Sigma^-$ and $ \Sigma^0$ productions and appearance of $\Sigma^+$, which is 
only produced in the final state interaction. 
This may be pointed out that in this paper we have not considered nuclear effect arising due to nucleon correlations.

SU(3) symmetry seems to work quite well in analyzing the semileptonic hyperon decays where symmetry
breaking effects are shown to be small in decay rates but play important role in explaining 
the observed asymmetries. There are some ambiguities in implementing the  
SU(3) symmetry in determining the weak form factors $f_2(q^2)$ and $g_3(q^2)$ in $|\Delta S| =1$ sector. 
SU(3) violating effects in the case of weak hyperon production and hyperon semileptonic decays are also 
related with the existence of second class currents in $|\Delta S| =1$ sector. 
The quasielastic production of hyperons induced by antineutrinos provides a unique opportunity to study these effects. 

The observation of quasielastic weak production of hyperons induced by antineutrinos and experimental 
data on total cross sections and differential cross section would provide very useful information
on weak form factors of nucleon-hyperon transition giving valuable information on various 
symmetry properties of weak hadronic currents like, SU(3) symmetry, $G$-invariance and CVC in 
$|\Delta S| =1$ sector. 

 \section{Acknowledgements}
 M. S. A. is thankful to Department of Science
and Technology(DST), Government of India for providing financial assistance under Grant No. SR/S2/HEP-18/2012.

\end{document}